# Medical Imaging with Deep Learning for COVID-19 Diagnosis: A Comprehensive Review

**Subrato Bharati [1], Prajoy Podder [2], M. Rubaiyat Hossain Mondal [3], V.B. Surya Prasath [4]**

[1,2,3] Institute of Information and Communication Technology, Bangladesh University of Engineering and Technology, Dhaka-1205, Bangladesh

[4] Division of Biomedical Informatics, Cincinnati Children's Hospital Medical Center, Cincinnati 45229, USA
[4] Department of Pediatrics, University of Cincinnati College of Medicine, Cincinnati, OH 45257 USA
[4] Department of Biomedical Informatics, College of Medicine, University of Cincinnati, Cincinnati, OH 45267 USA
[4] Department of Electrical Engineering and Computer Science, University of Cincinnati, OH 45221 USA
[1] subratobharati1@gmail.com, [2] prajoypodder@gmail.com, [3] rubaiyat97@iict.buet.ac.bd, [4] prasatsa@uc.edu

***Abstract***: **The outbreak of novel coronavirus disease (COVID-19) has claimed millions of lives and has affected all aspects of human life. This paper focuses on the application of deep learning (DL) models to medical imaging and drug discovery for managing COVID-19 disease. In this article, we detail various medical imaging-based studies such as X-rays and computed tomography (CT) images along with DL methods for classifying COVID-19 affected versus pneumonia. The applications of DL techniques to medical images are further described in terms of image localization, segmentation, registration, and classification leading to COVID-19 detection. The reviews of recent papers indicate that the highest classification accuracy of 99.80% is obtained when InstaCovNet-19 DL method is applied to an X-ray dataset of 361 COVID-19 patients, 362 pneumonia patients and 365 normal people. Furthermore, it can be seen that the best classification accuracy of 99.054% can be achieved when EDL_COVID DL method is applied to a CT image dataset of 7500 samples where COVID-19 patients, lung tumor patients and normal people are equal in number. Moreover, we illustrate the potential DL techniques in drug or vaccine discovery in combating the coronavirus. Finally, we address a number of problems, concerns and future research directions relevant to DL applications for COVID-19.**

## I. Introduction

The pandemic of novel coronavirus disease (COVID-19) and the ensuing containment efforts have created a global health pandemic affecting all spheres of human life [1-4]. At its beginning, with a small number of people impacted, it did not represent attacks of such immense magnitude when most of the cases were spontaneously resolved [5-8]. Coronavirus was declared a pandemic by the world health organization (WHO) in March 2020, with an incredibly unsafe for impacting the number of people in most of the countries, including those with worse health systems [9,10]. The virus is dangerous for two primary causes: it is novel, so researchers do not fully understand the virus, and it spreads by face to face or indirect communication through an infected human.

COVID-19 figures indicate the cause for a grave concern, with almost 153,598,296 patients infected worldwide and 3,218,287 patients evading the combat and surrendering to death as of May 03, 2021 [11]. The USA, as a global leader in healthcare development, has the largest number of COVID-19 casualties, followed by India, Brazil, French, and Turkey, among other nations [11].

Daily announcements of new outbreaks have been coming in at an unprecedented rate, leading governments and regulatory bodies worldwide to enforce a no-compromise lockdown to maintain social isolation and thereby contain the pandemic. In order to stem the spread of COVID-19, the global response has been swift and unequivocal, with the most affected countries barring their borders to transit and travel. The WHO has issued formal guidelines for individuals, governments, and national and multinational corporations to implement to ensure the disease's complete containment, thus breaking the chain that has culminated in this infectious disease [12]. The COVID-19 transmission is depicted in Fig.1.

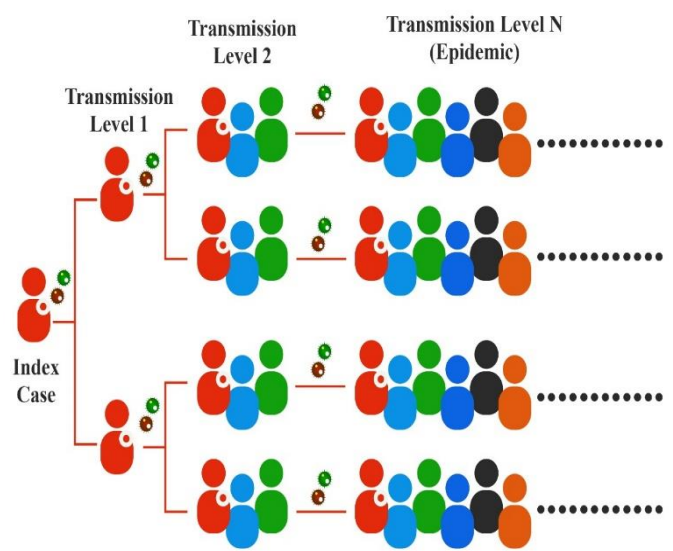

**Figure 1.** COVID-19 transmission human to human

In an urgent effort to fight the COVID-19 pandemic, studies have been undertaken in all guidelines, and deep learning (DL), together with image processing on medical images, has severely been investigated to pursue a complete cure [13,14]. Various research articles with related aims have been written [15-19]. The methods addressed are culled from a variety of peer-reviewed publications and hence contribute to creating a comprehensive collection of guidelines for the scientific community and even for regulatory authority tasked with combating the pandemic. One of the most significant challenges in COVID-19 study is a shortage of accurate and sufficient results. Owing to the limited number of studies done, numerous deaths and instances of virus infection go unreported. It is difficult to speculate about whether the failure factor for COVID-19 infection detection is three, three hundred, or perhaps more.

None of the countries across the world have provided accurate statistics on the virus's presence in a descriptive example of the overall population. However, development and research activities must continue, and therefore a fusion of knowledge is critical. The accessibility of high-quality, large-scale datasets in DL or machine learning (ML) schemes is critical to the precision of the findings. Information fusion allows several databases to be combined and used in DL models to improve prediction accuracy. Computed tomography (CT) photographs from different sources, for example, this fusion method has been combined as amount of various data fusion and served into DL and AI models [20]. The authors of [21] found similar knowledge fusion when radiological images of the chest from Cohen on GitHub repository were supplemented with Chest radiology images from the freely accessible in Kaggle. X-ray data in GitHub, RSNA, SIRM and Cohen dataset were linked and applied on CNN for detecting the infection of COVID-19 [22]. It is critical to recognize that the pandemic has reached a point where modern treatment services are overburdened. To accommodate the ever-growing population of patients, emergency rooms and intensive care centers have been extended beyond their normal capability. In such a situation, healthcare professionals and patient family members must make quick choices based on limited knowledge. The COVID-19 disease phenotype begins with moderate to no signs but quickly progresses to make patients highly critical, with sometimes catastrophic consequences due to the failures in multi-organ. The aim is to avoid such sudden impairments and diagnose infection as soon as possible using the small information base and available resources. The standard reverse transcription polymerase chain reaction (RT-PCR) test conducting a nose-throat-swab has low sensitivity and takes a long hour. Unavailability of RT-PCR and specialist laboratory equipment for conducting COVID detecting tests is unavoidable when patients' number is high. As a result, the need for tools that can help supplement capital is critical. Artificial intelligence (AI) and its subdivisions ML and DL are the tools. These tools can speed up to make a decision and increase patient-centered results. ML models have been produced in various studies to serve the same role with limited resources, resulting in accuracy (82–86%) and sensitivity (92–95%) that are substantially comparable to the RT-PCR test

[23]. Architecture such as DarkNet can classify and identify COVID-19 infections from radiology and assist the function in areas where the radiologists are limited. These limitations occur due to an excessive number of infected patients, which may be combined with radiological images to contribute to a more precise diagnosis of the infection at an initial stage, confirming the availability of healthcare [24].

Although several research works have been reported for the application of DL in the context of COVID-19, more research is required to make DL effective in managing COVID-19. This paper describes the necessary image processing and DL strategies suitable for the classification and detection of COVID-19. The rest of the paper is organized as follows. Section II provides an overview of COVID-19 and DL, while Section III reviews the DL techniques. Section IV discusses medical imaging along with DL methods. Comparative results are shown in Section V. The application of DL in drug discovery is described in Section VI. The existing challenges and prospects of DL are discussed in Section VII, followed by the concluding remarks in Section VIII.

## II. Overview of COVID-19 and DL

This section discusses the details of DL with COVID-19 and the application of DL to the processing and analysis of medical images in the current works.

### A. A brief of the COVID-19 pandemic and its current status

Many pneumonia cases were initially recorded in China [25]. The infection of COVID-19 patient was found in December 2019 in Wuhan, People's Republic of China (PRC). COVID-19 was declared and established to be a transmissible virus that was caused by severe acute respiratory syndrome coronavirus 2 (SARS-CoV-2). The researchers report that the virus was spread by wild bats [26]. This infectious virus is classified as a beta-coronavirus (beta-CoV), which also includes the SARS coronavirus. The authors of [27] observed that the SARS-CoV-2 outbreak began during the carnival of spring in China. The crowd of this massive mass from many nations catalyzed the infected virus's dissemination inside China and across borders of different countries. The WHO's China country office, Western Pacific Regional Office, and Headquarters have been operating diligently to consider the effect of COVID-19 beginning in the first week of January 2020 [28, 158, 159]. The WHO authorities declared the epidemic a PHEIC in the final week of January.

According to [29], COVID-19 spreads slowly in regions of elevated temperatures and humidity. Additionally, the writers emphasized the importance of steam therapy in decreasing the threat of COVID-19. Inhaled vapor travels from the pulmonary glands, assisting in the improvement of levels of oxygen. The authors of [30] concluded that the criteria governing the virus's development and spread are determined by environmental factors, sanitation, respiration water droplets, and physical interaction. On 3rd February 2020, the WHO released an effective approach to fight coronavirus [31], which includes recommendations and procedures for health professionals, physicians, and government leaders to tackle the COVID-19 outbreak while maintaining personal protection effectively. One strategy will be to test every potentially infected patient, separate them, and elicit information about their travels and contacts. One more critical technique recommended by the WHO is lockout, which can be a successful method of virus containment. The WHO's recommendation study [31], issued in March 2020, states that the spread of infection control was focused on experience obtained from case histories of patients who had SARS. Additionally, the study emphasizes the importance of observing such protocols, such as public use of surgical masks in COVID-19-affected environments, hand washing with hand wash or alcohol-based solutions.

Additionally, healthcare workers are not advised to contact affected patients. According to the WHO, there is no particular treatment suitable for COVID-19 at the moment, except for some best recommendations and practices [32]. The World Economic Forum has suggested using digital FDI to boost financial development in developed nations [33,34].

### B. Basics of DL

Neural networks and DL have attracted significant research interest [35-41]. Because of their potential to respond to many data forms in diverse contexts, these methods have been commonly utilized in applications, i.e., prediction and classification issues, smart houses, image recognition, object recognition, self-driving vehicles, etc. Various methods used in DL are depicted in Fig. 2. DL mimics the human brain's ability to process facts and make correct decisions. DL teaches a device to process inputs using various layers, similar to the human brain, to help data prediction and classification. These layers are similar to the layered filters deployed for NNs through the brain, in which every layer provides input to the next. The feedback loop is repeated before the same output is achieved. The accurate output is created by passing on weights to every layer, and these weights are changed through training to obtain the precise output. There are three types of DL techniques: supervised, unsupervised and semi-supervised. The DL model is trained where a proven pair of input-output is in supervised learning. Every detected rate serves as an input vector for the supervisory signal. It is the desired value. The approach predicts the labels of the target output by using current labels. Classification approaches use guided learning [42] and can be used to recognize traffic signals, faces, detect files spam, translate expression to text, and other scenarios. Semi-supervised learning is a methodology that bridges the gap between supervised and unsupervised ML techniques. Semi-supervised learning uses both classified and unlabeled values as training results. Semi-supervised learning is a form of learning that lies somewhere between unsupervised and supervised. When combined with a limited amount of labeled data, the unlabeled data will yield a significant increase in learning accuracy. Certain theoretical concepts about DL procedures keep on [43]. The first principle is that information in close proximity uses the same name. The second principle is the cluster assumption, which states that all data in a cluster have the same name. The third point is that rather than using the whole input space, the data is confined to a single dimension. Unsupervised learning entails determining the interrelationships between the data set's components and then classifying the data without the use of labels. Clustering, anomaly identification, and NN are some of the algorithms that use these strategies. The theory of clustering is to find common anomalies or elements in an information collection [44]. Unsupervised learning anomaly identification is commonly used in the domains of security [45,46]. For feature processing and extraction, most DL techniques use Artificial Neural Networks (ANN). The learning process employs a feedback strategy, in which every stage updates its input information in data to create a concise illustration [47]. The DL procedure refers to several layers needed to transform the information. During this transformation, a CAP is applied. The depth of CAP in a feed forward NN is determined by multiplying the number of hidden layers by the number of output layers. Since there could be a different signal that passes through several epochs in a layer in a Recurrent Neural Network (RNN), the depth of CAP could not be calculated [48]. CNN is the most often employed NN method for image manipulation [36,49, 50]. DL is the most reliable approach for image processing domains since the attribute extraction methodology is automatic and conducted during the training on the images in CNN. RNN functions in a similar way to CNN, with the exception that RNN is used to compute vocabulary. RNN employs feedback loops, in which the one layer in output is fed into the next input. RNN may be applied to time-series [51] video, financial details, document, audio, and other datasets. GANs (Generative Adversarial Networks) are focused on the generator network and discriminator concepts. The discriminator distinguishes between false and actual data, while the generator network generates fake data. GANs are more often found in applications that involve the production of images [52] from text, and these two networks act to improve the training method. The inception block in the Inception network of Google is used to calculate pooling and convolutions operations that operate in parallel for efficient managing of complex works. This is a higher degree of DL system that is applied to automate the works related to image processing [53].

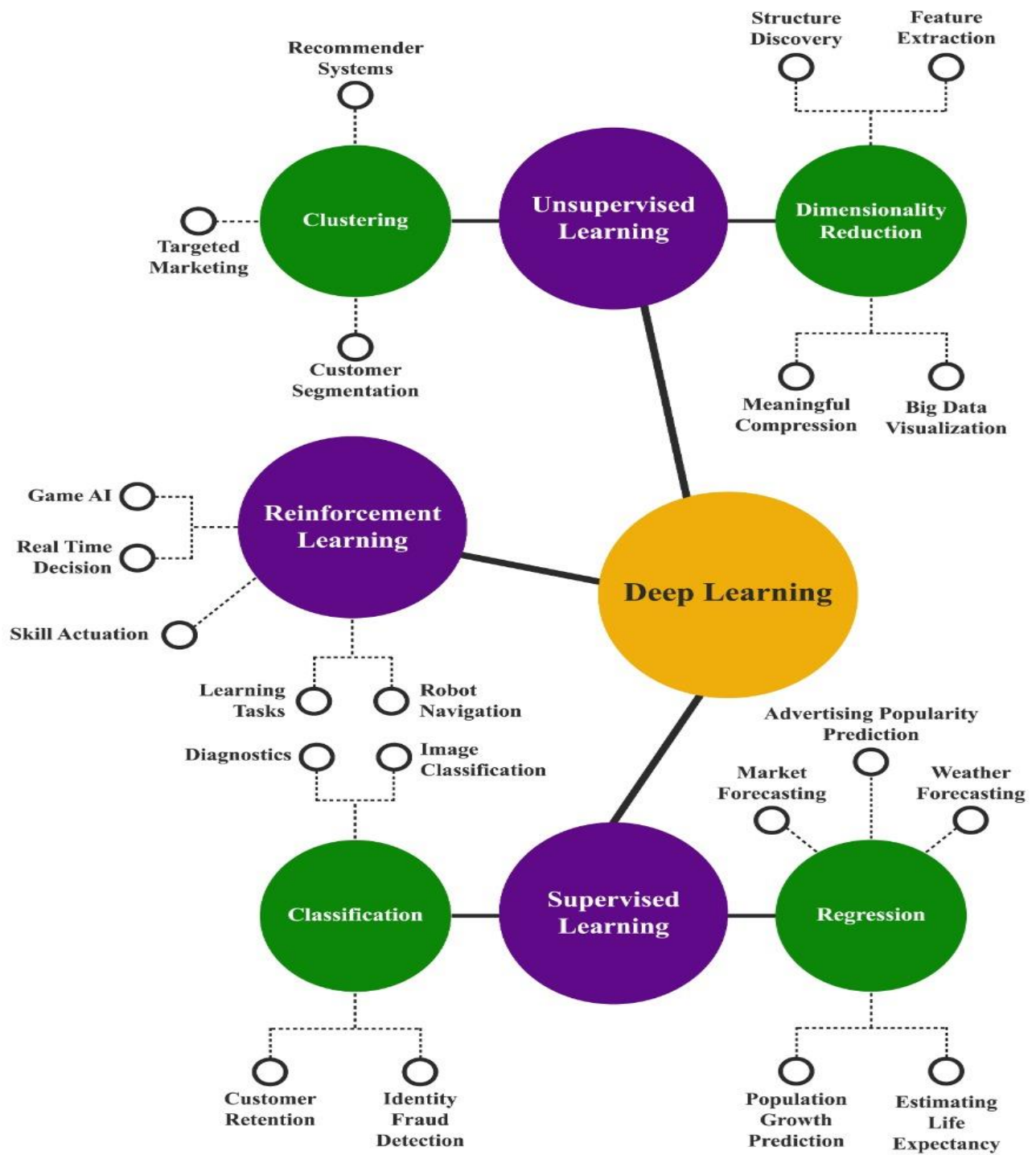

**Figure 2.** Different methods with DL applications

## III. A Review of DL Technologies of medical image processing

For the past three decades, advances in medical research have transformed clinical care, enabling physicians to more accurately diagnose and cure diseases [54]. However, physicians, like any other human being, are vulnerable to mistakes. The academic qualifications of a physician are determined not only by the character's intelligence degree, but rather by how they handle patients' conditions and the associated form of healthcare scheme that serves them [55]. This mix accounts for such a large variance in health results, and ML is the optimal approach for strengthening a physician's diagnostic and treatment abilities [56]. The success of ML procedures is contingent upon the features derived and the data representation used. ML algorithms face two primary challenges: reliability in searching all datasets with high-dimensional and teaching the architecture of ML to choose the best suitable role [57,58]. DL method has become one of the most widely employed methods for disease prediction and identification because it ensures a higher degree of precision. As shown in Fig. 3, the use of DL methods has resulted in novel inventions in the healthcare section. In the modern world, various forms of medical data have supplied doctors with large amounts of knowledge, such as CT, positron emission tomography (PET), X-ray and Magnetic resonance imaging (MRI) [59-61]. CNN is a widely used algorithm in the field of image processing and interpretation [49]. The work of [62] conducted a study of numerous DL approaches for image processing on medical images and offered the usage of DL procedure in segmentation, image categorization, and pattern recognition, among other applications. DL for image recognition is used in a variety of medical fields, including cancer diagnosis, ophthalmology, psychotherapy, cardiology, and neurology. Additionally, the researchers identified unsolved research issues in DL related to image processing. In the new era of electronic documents maintained by medical stakeholders and patients, AI has helped in simplifying medical image production. In [63], the authors discussed different AI procedures that could be used for image processing on medical images. The authors discovered that CNN, as well as big data modeling methods, have been extensively utilized in this research. Additionally, the authors highlighted the primary issues associated with the lack of high-quality labeled information for improved analysis.

### A. Classification

Classification of diseases is critical in the production of medical images. Several images are used as input samples during the classification processing step, and a particular diagnostic factor is produced as a DL systems output [64]. The authors of [65] applied DL to recognize lung nodules in 1995. The identification technique entails the use of 55-chest X-rays and 2-hidden layers in deep-neural. The radiologist detected 82 percent of nodules in the lung using this procedure. The authors of [66] applied multi-scale DL methods to detect nodules in lung deploying CT scans. Three hidden layers form the experimentation phase, which gets CT images as images in input and react to the output layer of the nodule in the lung. The work of [67] developed a CheXNet model. It has 121 convolutional layers and used 112120 chest X-rays as input to identify 14 various types of repository diseases. The researcher concludes from this analysis that the CheXNet algorithm outperforms the F1-metric performance range. The authors of [68] created a model for Diabetic Retinopathy diagnosis by training a ten-layer CNN. It has three fully interconnected layers on approximately 90000 images. On 5000 test images, experimental experiments achieve a sensitivity of 95% and a precision of 75%. Another research, [69], used IDx-DR-version-X2.1. It had to train DR images on 1.2 million samples for the purpose of defining DR. The results show that the built environment will improve sensitivity by 97 percent and specificity by 30%. The work of [70] suggested a CNN with multi-layer for skin lesion classification. A number of high-resolution images are deployed to train the multi-layer CNN. The suggested model obtains higher accuracy than another current model, based on data from a publicly accessible dataset of skin lesions.

### B. Localization

The photographs are fed to CNN for classification, at which point the image's content is exposed. Following image classification, the next stage of disease identification is image localization. It is accountable for defining the output position in the bounding box. This is referred to as localization with classification. Localization is a term that applies to determining the infection in the images. The anatomy localization is a critical pre-processing step in the medical diagnosis process since it allows the radiologist to identify these critical features. Numerous studies have been performed in recent years to localize disease using DL models. For instance, the authors of [71] offered a deep CNN-based architecture for classifying organs or body sections. The CNN was focused on 1675 patients with their 4298 X-rays in order to recognize the body's five organs: neck, legs, liver, lungs, and abdomen. The experimental findings indicate a grouping error of 5.9 percent and an AUC of 99.8%. Another ongoing study [72] developed a model for the location of T2 MR prostate photographs. Accurate prostate localization is complicated by a variety of factors, including thickness differences and irregular presentation. The authors trained images of prostate using Stacked Sparse Auto-Encoder (SSAE) for identification the diseases in prostates. The work was conducted on a dataset comprising MRI of 66 prostates, and the results are optimistic, indicating that the model outperforms current models. The researchers of [73] applied a DL technique to train DCE-MRI of 78 images of patients' kidneys and livers. Three distinct databases were analyzed to determine the location of disease in multi-organ. The suggested model as a whole is more precise in terms of disease localization in heterogeneous organs. Payer et al., [74] applied 3D CNN for land marking in medical images., the Spatial Configuration-Net (SCN) algorithm was used to combine correct response with the localization of landmark Experiments with 3D picture datasets utilizing CNN and the SCN architecture demonstrate increased precision. The work of [75] established a model. It aids in the localization of the fetal in a picture. These experiments obtained a precision of 69%, a recall of 80%, and an accuracy of 81%.

### C. Detection

Developing reliable ML models localizing, classifying, and identifying several items in a particular image has continued a central problem in the research of computer vision [76]. With modern advances in DL and the models in computer vision, developing medical image recognition systems has never been easier. Object recognition is a procedure that enables the identification and localization of various artifacts contained within a video or image. Item recognition is a method of computer vision. It is applied to recognize real-time objects with their instances. To find and recognize artifacts, object detection procedures either the prediction of train models or apply corresponding templates. Algorithms in object recognition make use of derived attributes and ML algorithms to distinguish instances of a certain object category. Object identification is a critical component in video monitoring, image processing systems, and applications in medical diagnostics [77]. For target tracking, the authors of [78] suggested the DL model with its Marginal Space. The used training patterns are applied to improve the efficiency of Deep Neural Network layers. The boundary delineation and estimated location, combined with a DL architecture, allow the segmentation of an image's outline. The experimental system consists of 869 patients and 2891 aortic valve videos, and it performs 45.2 percent better than previous ones. Another research [79] developed a new model for detecting interstitial lung disease and lymph nodes by training CNN utilizing threefold cross-validation. The experimental findings obtain an area under a receiver operating characteristic curve (AUC) of 0.95, which corresponds to an 85 percent sensitivity. Due to the memory constraints of the server, improving 2-D image recognition to 3-D image detection is a significant obstacle in the recognition of the image. The work [80] applied two-dimensional area application networks to catch explanations in objects of two-dimensional and then used a new method to integrate the two-dimensional solutions into three-dimensional solutions. The authors [81] published another intriguing paper in which they suggested a new model for detecting lung cancer. This procedure consists of two stages. The first phase employs a 3-D neural network to identify suspicious pulmonary nodules. The second stage involves the collection of the five finest nodules. The findings indicated a precision of 85.96% on training data and 81.42% on test data. The developers of [82] suggested a DL algorithm named the SSAE for detecting breast nuclei from images. SSAE is learned to extract high-level characteristics, which are then deployed as inputs. Next is to the classifier of softmax to recognize nuclei. The findings of this study indicated that the suggested model obtained an F-Measurement accuracy of 84.49 percent and a precision of 78.83 percent. The researchers of [83] deployed SSAE with CNN to achieve high-level functionality in a similar manner. A classifier like Softmax was applied to diagnose skin cancer. The experiment used 1417 photographs and obtained a precision of 91.4 percent and an F-measure of 89.4 percent.

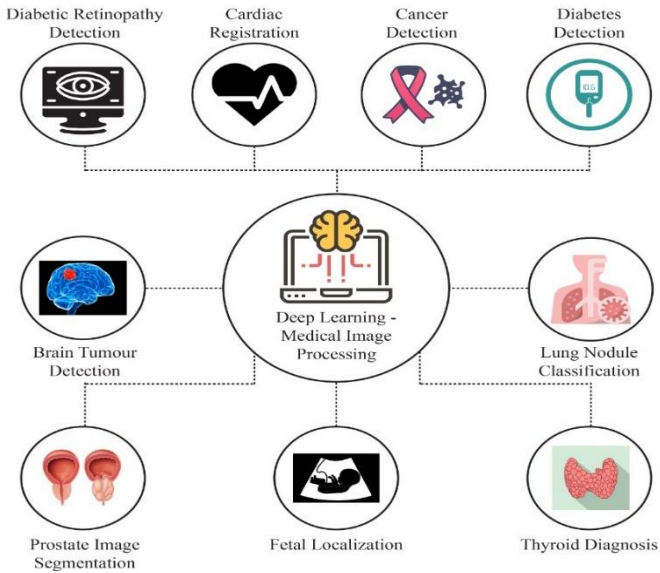

**Figure 3.** Image processing with DL applications

### D. Segmentation

In medical image processing, image segmentation is critical for disease detection [84]. Segmentation of an image is the process of dividing a visual image into several pieces. The aim of medical image segmentation is to make examining visual photos easier and more convenient. A series of medical segments comprising the whole medical image is the product of segmentation in a medical image [85]. Several inter-disciplinary approaches are now being utilized to analyze patient evidence in order to improve diagnostic precision. In [86], the authors suggested a controlled ANN methodology that incorporates cross-modality at all stages of the ANN. Furthermore, the authors develop an image segmentation approach based on CNN to represent soft tissue lesions from photographs collected using multiple inter-disciplinary procedures, i.e., CT, MRI, PET. Before using image processing, proper organ feature mining and analysis are needed. This may be useful in situations where the pictures are unclear or incorrect due to device failures. The shortage of synchronization between the targets in training and the performance dependencies is a major problem in current image processing methods. The authors of [87] suggested a new approach to address this problem by applying a basic learning process that embeds prior information into CNNs utilizing a thoroughly qualified regularization model. The detection of biomarkers in medical photographs aids diagnosis. The controlled DL is often applied for image processing, but it may be inaccurate due to the fact that it requires detailed awareness of the organ's location and characteristics. Detecting anomalous regions may help in the biomarker discovery phase. This anomaly detection may reveal details about the anatomical structure that was previously unknown. The researchers of [88] apply this concept to develop a DL-based Bayesian system, adopting that epistemic suspicions are related to the training dataset with its anatomical eccentricities. Using various ANN approaches, the researchers also apply Bayesian-based U-Net to practice on a development with poor characterizes of provided anatomy. Prostate MR scans with segmentation will reveal useful details for the diagnosis of prostate cancer. However, there are a number of drawbacks of utilizing MR scan segmentation, including a missed boundary area between the other organ and prostate, a multi-layered history, and a variation in prostate features. By applying boundary-weighted segmentation failure, the authors of [89] develop a new method named BOWDA-Net, which makes segmentation with its boundary detection easier. To solve the problem with limited image datasets, the authors use a boundary-weighted transfer learning technique.

### E. Registration

The process of transforming several datasets into a single coordinate model is known as image registration. In the fields of biological imaging and medical imaging, registration of an image is critical. To interpret or merge information from numerous medical sources, one must first register. A medical technician is typically instructed to view multiple photographs in various directions to minimize perceived overlap between them [90].

Radiation therapy is used by many cancer patients, creating it a popular cancer therapy [91-95]. Medical image recognition can assist many physicians and patients as the number of patients grows. Elastix is a completely automatic 3D deformable registration program that can be used in radiation oncology departments all over the world to improve radiotherapy procedures. The proposed model offers a global function quest that is computationally effective. It was previously challenging to find 3D registration software that could correctly handle improvements in orientation, localization, and strength variations. The suggested model saves physicians huge time and provides better registration results with medical confidence. Patients can have fewer repeat tests, conserve time, and be exposed to less radiation as a result of the improved capacity to integrate diagnostic scans. The effectiveness rate for the prostate was 97 percent, 93 percent for the influential vesicles, and 87% for the lymph nodes, according to the experimental results of the Haukeland Medical Center cancer dataset. The authors of [96] suggested a multi-class classifier to improve the registration accuracy score. Two phases are highlighted in the suggested model. In the first step, the Bayesian model was used to create a patch-based mark fusion approach to remove multiple features from atlas classifiers. Better registration accuracy is reached in the second stage, which uses label info. The findings of the cardiac MRI imaging dataset yielded dice scores of 0.92, 0.82, and 0.89 for the ventricular cavity at left, myocardium and ventricular cavity at right, respectively. Chee et al. [97] developed the registration 3D-image using a model of Self-supervised learning. The aim of this model is to perform picture registration in a short amount of time in order to identify the internal areas of the brain and integrate data from various sources. The findings of the experiments on the axial vision of brain CT-scans showed a 100x quicker run time. Lv et al. [98] proposed a CNN image registration model for capturing abdominal photographs, with the aim of detecting motion-free images of abdominal. The tests were carried out on ten distinct patients using a 1.5T MRI-scanner. Furthermore, as opposed to other current methods, the DNN model assisted in achieving improved registration performance.

### F. Summary

We reviewed applications of DL in medical imaging in this section, including segmentation, identification, classification, and registration. We have witnessed dramatic improvements in healthcare over the last three years as a result of new ways to transform people's lives in today's landscape of developments in DL. DL has been used in the science domain to classify different diseases via the study of biological photographs. We observe that DL algorithms penetrate a wide variety of medical image processing applications. Previously, physicians invested much time reviewing records, while the majority of tasks were performed manually. DL has started to streamline the lengthy procedure by offering improved outcomes, facilities, and advanced treatment instruments. Applications of DL remodel the healthcare sector by opening up fresh opportunities and improving people's medical lives.

## IV. Medical imaging with DL for COVID-19

This segment examines the capacity of DL in processing the medical image for combating the COVID-19 pandemic by the use of four strategies. As shown in Fig. 4, the techniques include disease detection, infection identification, coronavirus detection and care, vaccine, and drug development. X-rays are used to detect influenza and cancer in the earliest stages. However, CT scanning is a more advanced method that can spot minute variations in the anatomy of internal organs with the application of X-ray and computer vision technologies. Though X-rays are ineffective at diagnosing soft tissue problems due to their two-dimensional nature, CT scans use three-dimensional computer vision technologies to acquire different photographs from numerous perspectives of the organ in the body. Though all CT scans and X-rays provide internal body images with several components, the images produced by conventional X-rays often overlap. For instance, the ribs mask the heart and lungs, preventing a more precise diagnosis. On the other hand, for a CT scan, these overlapping elements are totally eliminated, allowing for a clear view of the internal anatomy and a clear understanding of the health issue. COVID-19 infection is often diagnosed based on basic pneumonia signs, an examination of the records of patient's travel, and proximity to new COVID-19 cases. However, chest imaging is critical in determining the degree of infection and subsequent treatment needs. COVID-19 instances are usually defined by patchy. The CT-scans served as markers of bilateral lung participation. Patients with chronic illnesses admitted to the intensive care unit (ICU) had a consolidative trend, while patients in non-ICU had a pattern of ground glass. In contrast, chest scans from patients with SARS and MERS disease showed unilateral signs. However, in 15% of cases, the original chest CT and X-ray of people still afflicted with the disease showed normalcy.

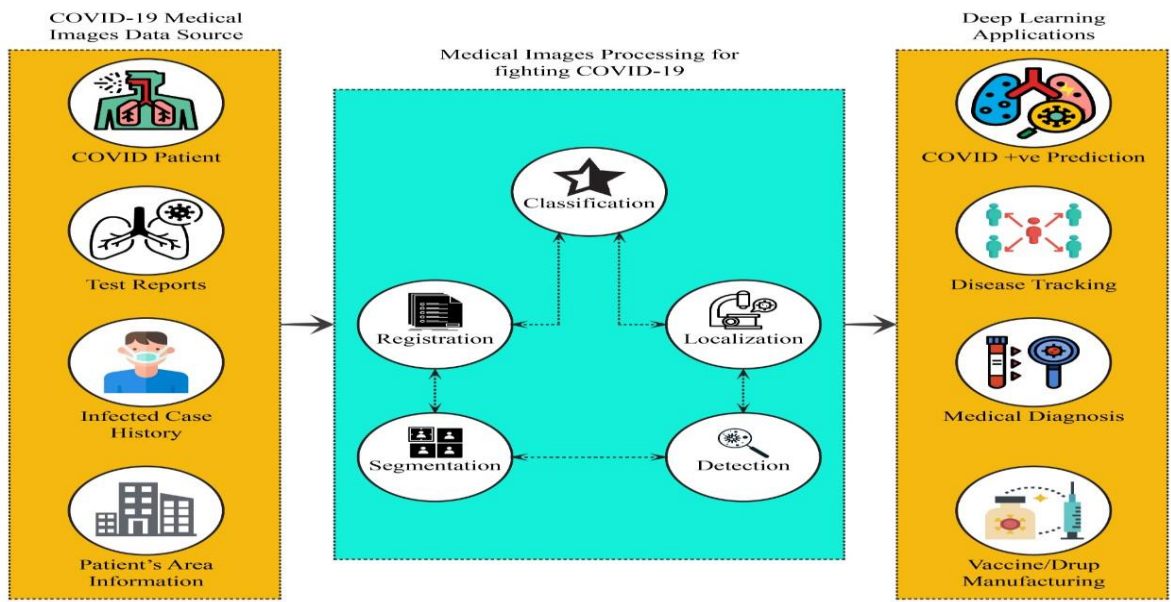

**Figure 4.** Image processing and DL for fighting COVID-19

This highlights the importance of additional validation through physical examinations or the application of DL-based methods, as mentioned in [99]. CT-scans are created as an X-ray rotates and takes images of a certain segment from a variety of angles. This radiology's are saved to the computer and processed further to produce a fresh picture that is free of conflicting elements. These photographs assist physicians in better comprehending internal systems by providing a full picture of their scale, composition, density, texture, and form. As a result, CT scans are known to be more powerful imaging tools than X-rays. The chest CT or X-ray is more often used to diagnose the existence of an illness, which may be caused by some other disorder. Additionally, the COVID-19 disorder is highly infectious, and imaging technology used on many patients is particularly dangerous. The scan machines are very complex pieces of equipment and washing them after each patient use is impractical. Moreover, if efforts are taken to disinfect, the likelihood of viruses exiting on the scan machine's surfaces is incredibly strong. Rather than imaging methods, swab tests have been shown to be more practical for COVID-19 diagnosis and identification. Numerous patients of COVID-19 initially have normal chest X-ray or CT results, but are subsequently shown to be positive for COVID-19. Although the conventional approach of RT-PCR is capable of accurately detecting disease, it has the disadvantage of requiring a longer detection period and additional reagents. The critical need is for rapid disease recognition using limited cost in the high populated countries. ML algorithms focused on image recognition play a critical contribution in reducing the overcoming crowd involved in performing the test in swab and other pertinent tasks. Fig. 5 illustrates sample X-ray and CT scan images of the patients of COVID-19.

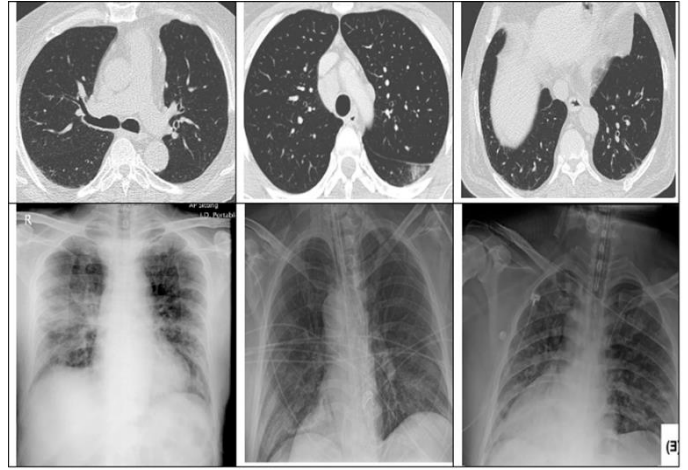

**Figure 5.** Upper is the example of CT-scans and lower is the example of X-rays of COVID-19 patients

## V. Comparative Studies

This section discusses the comparative results of the application of DL in the context of COVID-19. Table 1 illustrates some DL models with its result for COVID-19 and other diseases obtained from X-rays. Table 1 describes the splitting of training and testing samples of the imaging datasets and provides results of the DL algorithms applied to the dataset. The results are presented in terms of classification accuracy, precision, recall and F1-score. It can be seen that the highest classification accuracy of 99.80% is obtained when InstaCovNet-19 method [106] is applied to a X-ray dataset of 361 COVID-19 patients, 362 pneumonia patients and 365 normal people. Moreover, Table 2 also explains some methods of DL with COVID-19 detention applying chest CT-scans. Table 2 indicates that the best classification accuracy of 99.054% can be achieved when EDL_COVID [111] is applied to a CT image dataset of 7500 samples where COVID-19 patients, lung tumor patients and normal people are equal in number.

| Ref. | Images | Data Split | Method | Accuracy | Precision | Recall | F1-Score |
|------|--------|------------|--------|----------|-----------|--------|----------|
| [100] | COVID-19: 180 Normal: 200 | Training:75% Testing:25% | VGG16(Fine Tuning) | 85.26% | - | - | - |
| | | | ResNet18(Fine Tuning) | 88.42% | - | - | - |
| | | | ResNet101(Fine Tuning) | 87.37% | - | - | - |
| | | | ResNet50(Fine Tuning) | 92.63% | - | - | - |
| | | | VGG19(Fine Tuning) | 89.47% | - | - | - |
| [101] | Pneumonia(Bacteria): 330, Normal:310 COVID-19: 284, Pneumonia(Viral): 327 | 4-fold cross validation | 4-class CoroNet | 89.6% | 90% | 89.92% | 89.8% |
| | | | 3-class CoroNet | 95% | 95% | 96.9% | 95.6% |
| | | | Binary CoroNet | 99% | 98.3% | 99.3% | 98.5% |

| Ref. | Images | Data Split | Method | Accuracy | Precision | Recall | F1-Score |
|---|---|---|---|---|---|---|---|
| | 500 normal, 500 pneumonia and 157 COVID-19 | | 3-class CoroNet | 90.21% | 92% | 90% | 91% |
| [102] | 250 COVID-19, 2,753 pulmonary diseases and 3,520 healthy patients. | - | VGG 16 | 97% | - | 91.5% | 91.5% |
| [103] | COVID19: 85 Normal: 85 | Test: 120 Images Train: 50 Images | CNN | 94% | - | 100% | - |
| [104] | Normal: 8066, non-COVID19 pneumonia: 5538 COVID19: 358 | - | COVID-Net | 93.3% | 90.5%, 91.3%, 98.9% | 95.0%, 94.0%, 91.0% | |
| | | - | ResNet-50 | 90.6% | 88.2%, 86.8%, 98.8% | 97.0%, 92.0%, 83.0% | |
| | | - | VGG-19 | 83% | 83.1%, 75.0%, 98.4% | 98.0%, 90.0%, 58.7% | |
| [105] | COVID-19: 127, Pneumonia: 127, Normal: 127 | Training: 80% and testing: 20% | VGG 16 (Based on SVM) | 94.20% | - | 94.20% | 94.20% |
| | | | VGG 19 (Based on SVM) | 94.13% | - | 94.13% | 94.15% |
| | | | ResNet 18 (Based on SVM) | 94.26% | - | 94.26% | 94.25% |
| | | | ResNet 50 (Based on SVM) | 95.33% | - | 95.33% | 95.34% |
| | | | ResNet 101 (Based on SVM) | 93.53% | - | 93.53% | 93.54% |
| | | | InceptionV3 (Based on SVM) | 90.26% | - | 90.26% | 90.28% |
| | | | GoogleNet (Based on SVM) | 91.73% | - | 91.73% | 91.74% |
| [106] | COVID-19:361, Pneumonia:362 Normal:365 | Testing:20% and Training:80% | InstaCovNet-19 | 99.08% | 99% | 99.3% | 99% |
| | | | ResNet-101 | 98% | 98% | 98.3% | 98% |
| | | | Inception-v3 | 97% | 96.6% | 96.6% | 97% |
| | | | MobileNetV2 | 98% | 97.6% | 97.6% | 97.66% |
| | | | NASNet | 95% | 95.6% | 95.6% | 95.6% |
| | | | Xception | 97% | 97.3% | 97.3% | 97.% |
| [107] | COVID-19: 50, Normal: 50 | 5-fold cross-validation | AlexNet | 94.79% | 95.18% | 93.88% | 90.21% |
| | | | GoogleNet | 88.89% | 95.12% | 93.38% | 94.85% |
| | | | DenseNet201 | 90.56% | 97.85% | 92.92% | 86.08% |
| | | | ResNet18 | 87.28% | 95.91% | 89.13% | 94.07% |

| Ref. | Images | Data Split | Method | Accuracy | Precision | Recall | F1-Score |
|---|---|---|---|---|---|---|---|
| | | | ResNet101 | 97.18% | 98.64% | 98.64% | 97.05% |
| | | | ResNet50 | 94.55% | 98.28% | 88.19% | 90.69% |
| | | | Inceptionv3 | 96.40% | 89.28% | 89.25% | 94.11% |
| | | | VGG16 | 96.51% | 94.80% | 89.05% | 95.85% |
| | | | XceptionNet | 88.74% | 89.18% | 94.11% | 89.19% |
| | | | VGG19 | 88.86% | 96.58% | 88.63% | 91.04% |
| | | | Inceptionresnetv2 | 96.81% | 93.75% | 91.22% | 92.13% |
| [153] | Bacterial pneumonia: 2772, Normal: 2800, COVID-19: 341, Viral pneumonia: 1493 | 5-fold cross-validation | Inception-ResNetV2 | 95.3% | 84.0% | 70.7% | 76.8% |
| [24] | no-findings: 500, pneumonia: 500 | 5-fold cross-validation | DarkCovidNet | 98.08% | 98.03% | 95.13% | 96.51% |
| [16] | COVID-19: 203, healthy: 203 | Training: 70% and testing: 30% | ResNet-34 | 98.33% | 96.77% | 100.00% | 98.36% |
| [154] | Non-COVID: 5000, COVID-19: 184 | Train test split | SqueezeNet | - | - | 98% | - |
| [155] | COVID-19: 105, normal: 80, SARS: 11 | Training: 70% and testing: 30% | ResNet | 95.12% | - | 97.91% | - |
| | | | VGG19 | 97.35% | - | 98.23% | - |

*Table 1.* Comparative studies of X-ray images for COVID-19

| Ref. | Images | Data Split | Method | Accuracy (%) | Precision (%) | Recall (%) | F1-Score (%) |
|---|---|---|---|---|---|---|---|
| [108] | COVID-19 negative:740 and COVID-19 Positive:325 | Internal Validation 455 images for testing (COVID-19 negative: 360 and COVID-19 positive: 95) | M-Inception V3 | 89.5 | 71 | 88 | 77 |
| [109] | COVID-19: 777, Bacterial pneumonia: 505 | Training: 60%, Validation:10 %, and Testing:30% | VGG-16 | 84 | 80 | 89 | 84 |
| | | | DenseNet | 82 | 76 | 93 | 83 |
| | | | ResNet | 86 | 81 | 93 | 86 |
| | | | DRE-Net | 86 | 79 | 96 | 87 |
| | COVID-19: 777, Healthy: 708 | | VGG-16 | 90 | 92 | 89 | 91 |
| | | | DenseNet | 92 | 85 | 92 | 92 |
| | | | ResNet | 92 | 96 | 89 | 92 |

| Ref. | Images | Data Split | Method | Accuracy (%) | Precision (%) | Recall (%) | F1-Score (%) |
|------|--------|-----------|--------|--------------|---------------|------------|--------------|
| | | | DRE-Net | 94 | 96 | 93 | 94 |
| [110] | COVID-19: 349, Non-COVID-19: 463 | Training:80%, Validation: 10%, and Testing:10% | Ctnet-10 | 82.1 | - | - | - |
| | | | VGG-19 | 94.5 | - | - | - |
| [111] | COVID-19: 2500, Normal: 2500, Lung Tumor: 2500 | 5-Fold Cross Validation | EDL_COVID | 99.05 | - | 99.05 | 98.59 |
| | | | ResNet18_Soft max | 98.56 | - | 98.56 | 97.87 |
| | | | GoogLeNet_Sof tmax | 98.25 | - | 98.25 | 97.43 |
| | | | AlexNet_Softm ax | 98.16 | - | 98.16 | 97.3 |
| [112] | COVID-19: 15589, Normal:48260, | 5-Fold Cross Validation | Modified ResNet 50 V2 | 98.49 | - | 96.83 | - |
| | | | Xception | 96.55 | - | 97.25 | - |
| | | | ResNet 50 V2 | 97.52 | - | 97.74 | - |
| [113] | COVID-19: 349, Non-COVID-19: 397 | Training:60%, Validation: 15%, and Testing:25% | EfficientNetB0 | 82 | 84.7 | 82.2 | 81.5 |
| | | | EfficientNetB1 | 71 | 72.7 | 71.8 | 71.2 |
| | | | EfficientNetB2 | 77 | 76.8 | 76.8 | 76.8 |
| | | | EfficientNetB3 | 76 | 76.9 | 76.5 | 76.3 |
| | | | EfficientNetB4 | 79 | 79.1 | 78.9 | 78.8 |
| | | | EfficientNetB5 | 82 | 81.7 | 81.7 | 81.7 |
| | | | InceptionV3 | 82 | 82.5 | 81.4 | 81.5 |
| | | | NASNetLarge | 77 | 77.2 | 77.0 | 76.8 |
| | | | NASNetMobile | 76 | 75.9 | 75.7 | 75.7 |
| | | | SeResnet50 | 75 | 75.5 | 74.5 | 74.5 |
| | | | ResNext50 | 81 | 81.0 | 80.6 | 80.6 |
| [156] | COVID-19: 1029, Normal: 1387 | Train test split | Hybrid 3D | 90.8 | - | 84 | - |
| | COVID-19: 168, non-COVID-19: 168 | 10-Fold Cross Validation | Deep CNN | 96.28 | 94.81 | 97.92 | 96.34 |

*Table 2.* Comparative studies of CT images for COVID-19

## VI. Contribution of DL in Drug Discovery

In this part, we will look at some of the more recent advances in Deep Neural Networks (DNN), which can be used at various levels of drug development. Although traditional vaccine development methods are inefficient and time-consuming, they are often ineffective when there is a chance of failure because pathogens are difficult to produce under laboratory conditions, as well as an abundance of antigens with little promise of providing protection against the disease of interest. AI-based vaccine production, on the other hand, avoids these flaws and allows researchers to discover new antigen vaccine candidates [114]. Furthermore, ML technology may be used to develop and consolidate more intelligent supervision in supply chains [115]. When it comes to controlling COVID-19, using different research designs and methodological techniques to complete clinical studies in a statistically successful manner is critical [116]. Combining MD simulations with AI methods as an important part of a general pharmacological model of drug activity seems to be a valuable mechanism for developing safer and more efficacious drugs [117]. ML processes, like ANN, have a long tradition of use in medicine and engineering to forecast compound behaviors and other dynamic variables [118, 119]. In ANNs, there are a number of layers that work together to make one output layer the next layer input before the list of outputs is depleted and an output layer that can predict a property is found. To render ANNs amenable to matrix operations, a matrix that expresses the weights of relationships between the layers represents pairs of linked layers. ANNs, especially deep versions of these networks, have recently played a significant role in applications to discover new drugs [120]. An input layer, hidden layers, and an output layer are seen in Figure 6 of a traditional DL form. This framework can be used as an accurate indicator for drug development targets using a basic technique [121]. The most popular use of DL approaches is to solve problems in the behavior prediction phase [119]. D Land ANN principles were first implemented in the 1980s [122]. Deep Neural Networks (DNNs), which have a number of hidden layers and are a big interest of information technology firms, was created as a way to solve problems like speech recognition [123]. The triggering of a binary logic gate in the networks was analogized to imitate neuron excitation in the human brain in both ANN [124]. The quantitative structure-activity relationship (QSAR) technique is an example of this situation that is widely used for property prediction. For e.g., log P, solubility, and bioactivity for provided chemical structures are predicted in this scenario [125]. If the aim of QSAR ranking applications is to find compounds with higher activities, the optimal ranking goal is for compounds with higher activities to be ranked higher than compounds with lower activities [125]. If proper AI approaches are used, we will enter an age in which clinical trial errors are reduced to the lowest rate possible and the drug discovery phase can be done quicker, easier, and more effectively [126]. As a result, using ML methods, information can be extracted from six types of evidence: proteomics, microarrays, genomics, biological structures, data mining, and text mining [127]. Furthermore, when the training set includes a large number of details, advanced feature learning allows DL to achieve high accuracy in recognition [124]. If DL approaches are to be feasible and reliable, more specialized programming expertise and hardware technologies are needed due to the fast-paced and increasing time complexity caused by network design complexity [124]. These AI-related approaches produce results such as constructing a drug from scratch and determining the right structure based on experimental experiments, all of which can be accomplished using modeling and quantum chemistry [128]. The possibility of incorporating AI into the drug manufacturing phase is no longer a pipe dream [129]. AI-related computing algorithms have progressed to the point that computer-based inference engines will now draw previously unheard-of conclusions [130]. Understanding how COVID-19 knows the host cell is critical in the battle against COVID-19 as a deadly and life-threatening disease. Furthermore, without the necessary epidemiological evidence, it is difficult to accurately track and forecast how an outbreak spreads during an outbreak [131]. Other data sets, on the other hand, rely mostly on aggregated case counts in each regional area [132]. A Deep Boltzmann Machine (DBM) is a structured model that is built on or adapted from probability theory and consisting of many layers of random and often latent variables [133, 134]. DBMs have been used as successful drug exploration tools. Figure 7 depicts a DBM that has been seen in many drug development studies [135]. DBMs are a kind of generative model that can be used to perform feature learning. These classifiers are suitable for use as COVID-19 vaccine methods classifiers. This paradigm may also be used to provide a single representation that combines modalities. It is also useful in situations where human experience is minimal, such as hyperspectral imaging. DBM may also approximate the characteristics of prior knowledge base samples without any input from the labels [136], making it a better candidate than others for assisting scientists in determining the right route to produce vaccines [137]. A Deep Belief Network (DBN) is seen in Figure 8 and is used in many dynamic applications. DBM [138] is a highly effective hierarchical generative model for extracting features capable of representing highly variant functions and finding their manifold [139]. DBN has a benefit over conventional deep model training methods like multilayer perceptron in that it can depend on a specific unsupervised pre-training technique to avoid over-fitting to the training set through [138, 140]. In addition, Fig. 9 shows a Hopfield neural network (Hopfield net) structure, which is used for drug development and prediction problems. Hopfield nets are a form of recurrent ANN, and they are an important tool for solving optimization problems as well as a suitable technique for the subject type of address memories [141]. Despite the fact that it is an ancient method, it has been applied to a wide range of bioinformatics and biological issues [142, 143]. Restricted Boltzmann Machine (RBM) is a kind of generative stochastic ANN that can be trained using a probability distribution as inputs. As a result, it is a viable choice for solving drug development and bioinformatics problems [143-145]. An RBM's configuration is depicted in Figure 10. Sangari and Sethares [146] were the first to create RBM, a kind of conventional DBM that can be generated by linking DBM neurons in a hierarchical manner. Without any relations between hidden layers, RBMs have been successfully used to model distributions over binary-valued data [134]. When a single representation that fuses modalities together has to be extracted, the model comes in handy. Learning a joint density model over the space of multimodal inputs is the secret. While various methods for speeding up the drug discovery process have recently been proposed, we will attempt to review those approaches that play a critical role in AI-assisted drug discovery, especially in terms of applicability, efficiency, and generalizability. Since such methodologies must not only be realistic in order to conserve time, but they must also be versatile enough to integrate with other traditional approaches. Training each RBM to model samples from the posterior distribution of the previous RBM raises a variational lower bound on the DBM's probability, which can be used to initialize the joint model [147]. Harmonium [148] was the RBMS prototype created by Paul Smolensky in 1986. However, it was not until Geoffrey Hinton and his colleagues fitted them with rapid learning algorithms that they developed in the mid-2000s that it rose to popularity. Long Short-Term Memory (LSTM) is a form of RNN used in the field of DL [147]. Unlike other types of regular feedforward neural networks, this one has input relations between its layers. LSTM is a powerful ANN that can address a wide range of drug development problems [149-152]. Figure 11 depicts the arrangement of an LSTM.

# VII. Lessons learned, obstacles encountered, and potential directives

Although DL has grown in popularity and achieved remarkable results for basic 2D pictures, there are limitations to obtaining a comparable level of success in the processing of the medical image. The review in this area is still ongoing, but the following are some of the lessons learned [160]:

1) A significant impediment is the lack of massive databases of high-quality photos for instruction. In this scenario, data synthesis could be a viable option for integrating data obtained from disparate sources.

2) The bulk of recent DL models are implemented on two-dimensional imaging data. CT and MRI, on the other hand, are typically three-dimensional and thereby add another dimension to the current challenge. Due to the fact that traditional DL models are not adjusted for this, expertise is critical as DL models are applied to these videos.

3) One of the main problems of medical image analysis is the lack of standardization in the collection of images. It is critical to realize that as data diversity increases, the need for broader datasets increases to confirm that the DL algorithms produce robust explanations. The most effective method of resolving this problem is the use of transfer learning. It streamlines pre-processing and acquisition problems and avoids scanners.

### A. Obstacles and Concerns

The following are the problems and concerns associated with DL applications for medical image processing for the purpose of containing the COVID-19:

### 1) Safety
Obtaining images of high-quality COVID-19 and broader databases is a significant difficulty when contemplating the privacy of patient records.

### 2) Unpredictability in Outbreak Trend
Since the outbreak evidence follows a dynamic pattern and exhibits extreme variance in activity across many nations, the predictability of outbreaks becomes an additional difficulty.

### 3) Legislation and Transparency
Countries around the world have enacted stringent protocols governing the exchange of COVID-19 results. One of the main protocols expressly specifies that the bare minimum volume of specimens and data must be obtained from infected patients in the shortest time. As a result, analysis becomes more complex.

### B. Future directives
As shown and examined in previous research, there are several DL implementations applied to a variety of datasets using a variety of assessment standards, with radiology imaging datasets becoming particularly prevalent. However, the application of these methodologies in real-world patient practice scenarios is a significant challenge, necessitating the urgent need for benchmarking frameworks for evaluating and comparing emerging methodologies. These systems can facilitate the usage of analytical hardware infrastructures by taking into account similar medical information, data pre-processing processes, and assessment requirements for different AI methods that ensure data transparency and interpretability.

The people of the modern world are much luckier than their forefathers, who lived through the Spanish flu pandemic of 1918, and we have been lavished by sophisticated technologies. AI has been widely applied in all areas of human life. Given that AI has permeated all spheres of existence, the same technologies should be used and thoroughly investigated in order to tackle the COVID-19.

Without the advantages described previously, the deployment of AI will significantly aid in the eradication of false news on social networking sites. AI will be used to sort out details on government policy, pandemic control procedures, and the evidence behind viral dissemination and containment, meaning that only authentic information enters the general public, obviating any possibility of premature hysteria. At this point in the never-ending COVID-19.

An only point of light is the invention of a new COVID-19 vaccine. AI has an enormous possibility in this field, as it can be used to investigate the virus's genetic and protein structure, thus expediting the drug discovery phase.

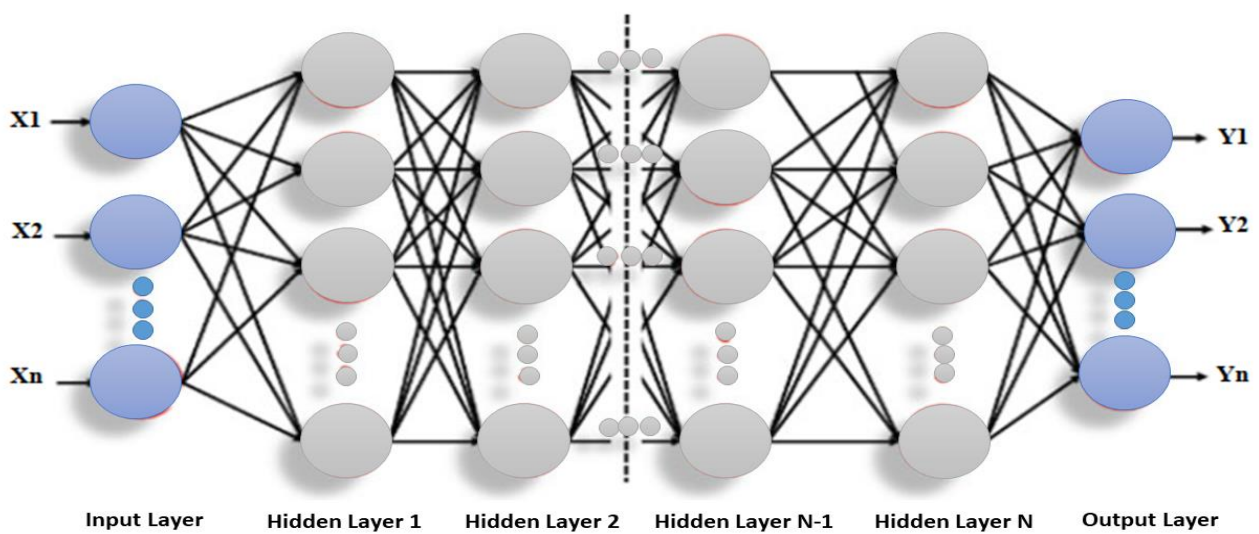

**Figure 6.** A model of Deep CNN for application in drug

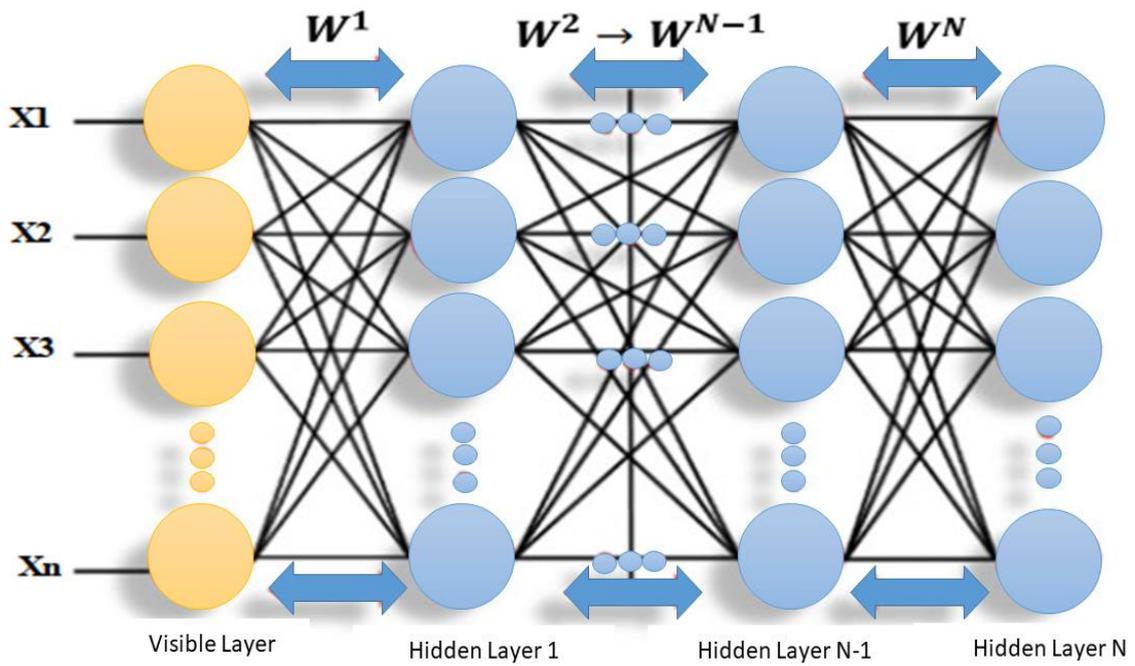

**Figure 7.** DBM structure [135]

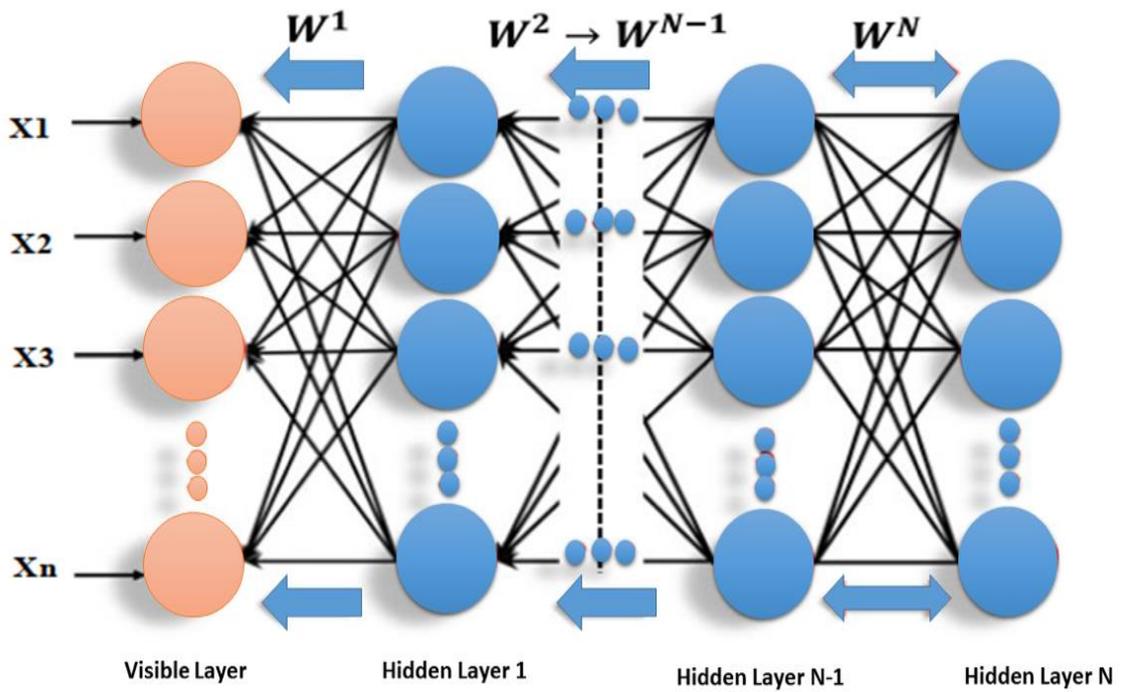

**Figure 8.** DBN structure [138]

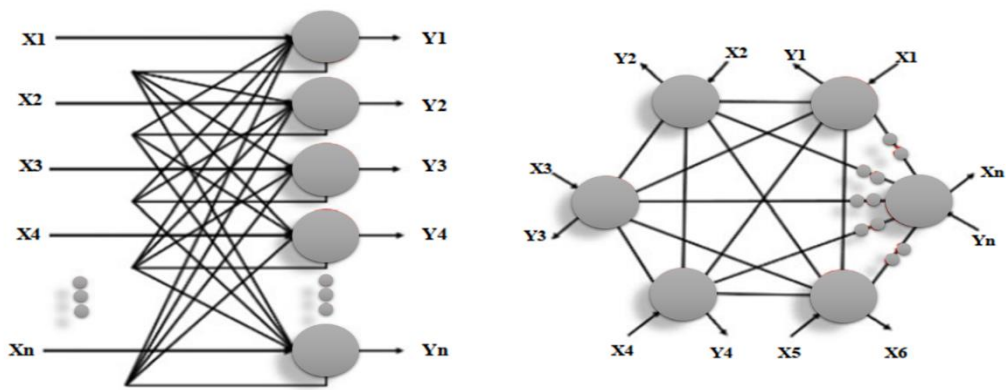

**Figure 9.** Configuration of Hopfield NN [141]

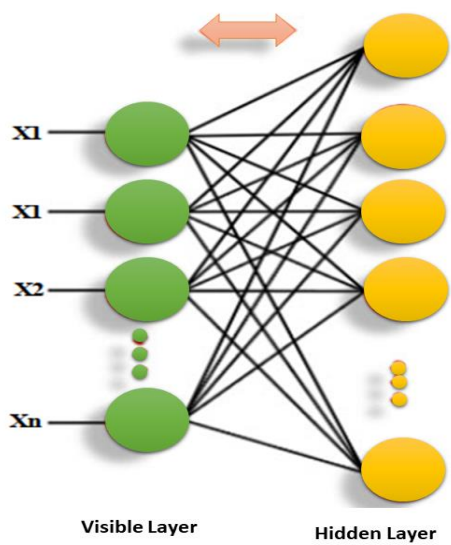

**Figure 10.** RBM model [146]

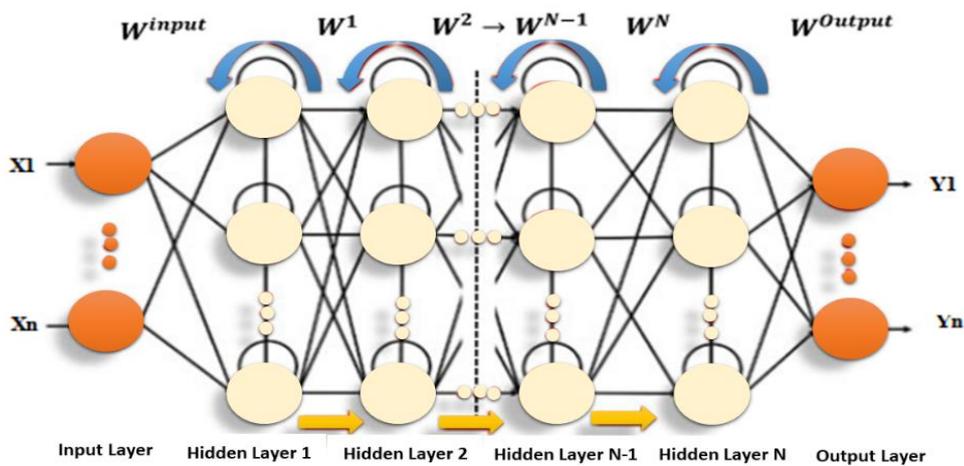

**Figure 11.** LSTM structure [147]

## VIII. Conclusion

Deep learning is an excellent method for providing an incredible response to combat against infected diseases like COVID-19. This paper outlines state-of-art efforts about managing COVID-19 pandemic and describes several implementations of DL for medical imaging. DL has been used to develop a variety of COVID-19 disturbance solutions, including disease detection, spread of virus monitoring, drug development, care and diagnosis, vaccine testing etc. Despite the positive outcomes, using DL to process COVID-19 medical images involve a significant amount of time and effort, as well as strong collaboration among government, industry, and academia. Data safety, outbreak trend heterogeneity, control and openness, and the differentiation between non-COVID-19 and infected COVID-19 symptoms are among the problems and concerns posed by previous studies. Finally, this paper presents few potential possible paths for DL implementations in medical imaging for COVID-19. With the integration of DL, image processing tools, mobile communications, computer science, and biomedicine and other research areas, COVID-19 will be better managed. In the fight against the COVID-19 pandemic, this study serves as a strong reference point and inspires new research on DL and medical imaging.

## References

[1] Solomon, M.D., McNulty, E.J., Rana, J.S., Leong, T.K., Lee, C., Sung, S.H., Ambrosy, A.P., Sidney, S. and Go, A.S., 2020, "The Covid-19 pandemic and the incidence of acute myocardial infarction," New England Journal of Medicine, 383(7), pp.691-693.

[2] Daniel, J., 2020, "Education and the COVID-19 pandemic," Prospects, 49(1), pp.91-96.

[3] Pfefferbaum, B. and North, C.S., 2020, "Mental health and the Covid-19 pandemic," New England Journal of Medicine, 383(6), pp.510-512.

[4] Spinelli, A., Pellino, G., 2020 Jun, "COVID-19 pandemic: perspectives on an unfolding crisis," Journal of British Surgery, 107(7):785-7.

[5] Vaishya R, Javaid M, Khan IH, Haleem A. Artificial Intelligence (AI) applications for COVID-19 pandemic. Diabetes & Metabolic Syndrome: Clinical Research & Reviews. 2020 Jul 1;14(4):337-9.

[6] Mondal, M.R.H, Bharati, S., Podder, P., Podder, P., 2020, "Data analytics for novel coronavirus disease," Informatics in Medicine Unlocked. 20(2020), 100374.

[7] Bharati, S., Podder, P., Mondal, M. R. H., & Gandhi, N., 2020, December, "Optimized NASNet for Diagnosis of COVID-19 from Lung CT Images," In International Conference on Intelligent Systems Design and Applications (pp. 647-656). Springer, Cham.

[8] Cevik M, Bamford C, Ho A. COVID-19 pandemic–a focused review for clinicians. Clinical Microbiology and Infection. 2020 Apr 25.

[9] Cucinotta D, Vanelli M. WHO declares COVID-19 a pandemic. Acta Bio Medica: Atenei Parmensis. 2020;91(1):157.

[10] Tosepu R, Gunawan J, Effendy DS, Lestari H, Bahar H, Asfian P. Correlation between weather and Covid-19 pandemic in Jakarta, Indonesia. Science of The Total Environment. 2020 Jul 10;725:138436.

[11] https://www.worldometers.info/coronavirus/ [last access 3 May 2021]

[12] Mossa-Basha M, Medverd J, Linnau KF, Lynch JB, Wener MH, Kicska G, Staiger T, Sahani DV. Policies and guidelines for COVID-19 preparedness: experiences from the University of Washington. Radiology. 2020 Aug;296(2):E26-31.

[13] Hakak S, Khan WZ, Imran M, Choo KK, Shoaib M. Have you been a victim of COVID-19-related cyber incidents? Survey, taxonomy, and mitigation strategies. IEEE Access. 2020 Jun 30;8:124134-44.

[14] Iwendi C, Bashir AK, Peshkar A, Sujatha R, Chatterjee JM, Pasupuleti S, Mishra R, Pillai S, Jo O. COVID-19 patient health prediction using boosted random forest algorithm. Frontiers in public health. 2020 Jul 3;8:357.

[15] Podder, P., Khamparia, A., Mondal, M., Rahman, M.A. and Bharati, S., 2021, "Forecasting the Spread of COVID-19 and ICU Requirements," International Journal of Online & Biomedical Engineering, 17(5), pp. 81-99.

[16] Nayak, S.R., Nayak, D.R., Sinha, U., Arora, V. and Pachori, R.B., 2021, "Application of deep learning techniques for detection of COVID-19 cases using chest X-ray images: A comprehensive study," Biomedical Signal Processing and Control, 64, p.102365.

[17] Horry, M.J., Chakraborty, S., Paul, M., Ulhaq, A., Pradhan, B., Saha, M. and Shukla, N., 2020, "COVID-19 detection through transfer learning using multimodal imaging data," IEEE Access, 8, pp.149808-149824.

[18] Elaziz MA, Hosny KM, Salah A, Darwish MM, Lu S, Sahlol AT. New machine learning method for image-based diagnosis of COVID-19. Plos one. 2020 Jun 26;15(6):e0235187.

[19] Afshar P, Heidarian S, Naderkhani F, Oikonomou A, Plataniotis KN, Mohammadi A. Covid-caps: A capsule network-based framework for identification of covid-19 cases from x-ray images. Pattern Recognition Letters. 2020 Oct 1;138:638-43.

[20] Wang S, Kang B, Ma J, Zeng X, Xiao M, Guo J, Cai M, Yang J, Li Y, Meng X, Xu B. A deep learning algorithm using CT images to screen for Corona Virus Disease (COVID-19). European Radiology. 2021 Feb 24:1-9.

[21] Ghoshal B, Tucker A. Estimating uncertainty and interpretability in deep learning for coronavirus (COVID-19) detection. arXiv preprint arXiv:2003.10769. 2020 Mar 22.

[22] Apostolopoulos, I.D. and Mpesiana, T.A., 2020, "Covid-19: automatic detection from x-ray images utilizing transfer learning with convolutional neural networks. Physical and Engineering Sciences in Medicine," 43(2), pp.635-640.

[23] Brinati, D., Campagner, A., Ferrari, D., Locatelli, M., Banfi, G. and Cabitza, F., 2020, "Detection of COVID-19 infection from routine blood exams with machine learning: a feasibility study," Journal of medical systems, 44(8), pp.1-12.

[24] Ozturk, T., Talo, M., Yildirim, E.A., Baloglu, U.B., Yildirim, O. and Acharya, U.R., 2020, "Automated detection of COVID-19 cases using deep neural networks with X-ray images," Computers in biology and medicine, 121, p.103792.

[25] Pham, Q.V., Nguyen, D.C., Huynh-The, T., Hwang, W.J. and Pathirana, P.N., 2020, "Artificial Intelligence (AI) and Big Data for Coronavirus (COVID-19) Pandemic: A Survey on the State-of-the-Arts," IEEE Access, 8, pp.130820-130839.

[26] Andersen, K.G., Rambaut, A., Lipkin, W.I., Holmes, E.C. and Garry, R.F., 2020, "The proximal origin of SARS-CoV-2," Nature medicine, 26(4), pp.450-452.

[27] Unhale, S.S., Ansar, Q.B., Sanap, S., Thakhre, S., Wadatkar, S., Bairagi, R., Sagrule, S. and Biyani, K.R., 2020, "A review on corona virus (COVID-19)," World Journal of Pharmaceutical and Life Sciences, 6(4), pp.109-115.

[28] WHO. (2019). Coronavirus disease (COVID-19) outbreak https://www.who.int/westernpacific/emergencies/covid-19. [Last Access 3 May 2021].

[29] Kampf G, Todt D, Pfaender S, Steinmann E. Persistence of coronaviruses on inanimate surfaces and their inactivation with biocidal agents. Journal of hospital infection. 2020 Mar 1;104(3):246-51.

[30] Shereen MA, Khan S, Kazmi A, Bashir N, Siddique R. COVID-19 infection: Origin, transmission, and characteristics of human coronaviruses. Journal of advanced research. 2020 Jul 1;24:91-8.

[31] WHO coronavirus disease (COVID-19) dashboard. (2020) Accessed 15.05.20 https://covid19.who.int/.

[32] Coronavirus (COVID-19): News, analysis and resources. (2020) Accessed 27.04.20 https://unctad.org/en/Pages/coronavirus.aspx.

[33] How digital investment can help the COVID-19 recovery. (2020) Accessed 28.04.20 https://www.weforum.org/agenda/2020/04/covid-19-digital-foreign-direct-investment-economic-recovery/.

[34] Bharati S. How Artificial Intelligence Impacts Businesses in the Period of Pandemics?. Journal of the International Academy for Case Studies. 2020 Jul 1;26(5):1-2.

[35] Chouhan, V., Singh, S.K., Khamparia, A., Gupta, D., Tiwari, P., Moreira, C., Damaševičius, R. and De Albuquerque, V.H.C., 2020, "A novel transfer learning based approach for pneumonia detection in chest X-ray images," Applied Sciences, 10(2), p.559.

[36] Khamparia, A., Singh, A., Anand, D., Gupta, D., Khanna, A., Kumar, N.A. and Tan, J., 2018, "A novel deep learning-based multi-model ensemble method for the prediction of neuromuscular disorders," Neural computing and applications, pp.1-13.

[37] Thanh, D.N., Hai, N.H., Tiwari, P. and Prasath, V.S., 2021, "Skin lesion segmentation method for dermoscopic images with convolutional neural networks and semantic segmentation," Computer Optics, 45(1).

[38] Hakak S, Khan WZ, Imran M, Choo KK, Shoaib M. Have you been a victim of COVID-19-related cyber incidents? Survey, taxonomy, and mitigation strategies. IEEE Access. 2020 Jun 30;8:124134-44.

[39] Bharati, S., Podder, P. and Mondal, M.R.H., 2020, "Hybrid deep learning for detecting lung diseases from X-ray images," Informatics in Medicine Unlocked, 20, p.100391.

[40] Bharati, S., Podder, P. and Mondal, M.R.H., 2020, "Artificial neural network based breast cancer screening: a comprehensive review," International Journal of Computer Information Systems and Industrial Management Applications., 12, pp.125-137.

[41] Bharati, S. and Podder, P., 2021, "1 Performance of CNN for predicting cancerous lung nodules using LightGBM," In Artificial Intelligence for Data-Driven Medical Diagnosis, De Gruyter, pp. 1-18.

[42] Patel, H., Singh Rajput, D., Thippa Reddy, G., Iwendi, C., Kashif Bashir, A. and Jo, O., 2020, "A review on classification of imbalanced data for wireless sensor networks," International Journal of Distributed Sensor Networks, 16(4), p.1550147720916404.

[43] Cheng Y. Semi-supervised learning for neural machine translation. In Joint training for neural machine translation 2019 (pp. 25-40). Springer, Singapore.

[44] De Simone A, Jacques T. Guiding new physics searches with unsupervised learning. The European Physical Journal C. 2019 Apr;79(4):1-5.

[45] Lima AQ, Keegan B. Challenges of using machine learning algorithms for cybersecurity: a study of threat-classification models applied to social media communication data. In Cyber influence and cognitive threats 2020 Jan 1 (pp. 33-52). Academic Press.

[46] Bharati, S., Podder, P., Mondal, M.R.H., Robel, M.R.A., 2020, "Threats and countermeasures of cyber security in direct and remote vehicle communication systems", Journal of Information Assurance and Security, 15(4).

[47] Shanmuganathan S. Artificial neural network modelling: An introduction. In Artificial neural network modelling 2016 (pp. 1-14). Springer, Cham.

[48] Yu Y, Si X, Hu C, Zhang J. A review of recurrent neural networks: LSTM cells and network architectures. Neural computation. 2019 Jul;31(7):1235-70.

[49] Huynh-The T, Hua CH, Pham QV, Kim DS. MCNet: An efficient CNN architecture for robust automatic modulation classification. IEEE Communications Letters. 2020 Jan 20;24(4):811-5.

[50] Rawat W, Wang Z. Deep convolutional neural networks for image classification: A comprehensive review. Neural computation. 2017 Sep;29(9):2352-449.

[51] Che Z, Purushotham S, Cho K, Sontag D, Liu Y. Recurrent neural networks for multivariate time series with missing values. Scientific reports. 2018 Apr 17;8(1):1-2.

[52] Greenspan H, Van Ginneken B, Summers RM. Guest editorial deep learning in medical imaging: Overview and future promise of an exciting new technique. IEEE Transactions on Medical Imaging. 2016 Apr 29;35(5):1153-9.

[53] Alom MZ, Yakopcic C, Nasrin MS, Taha TM, Asari VK. Breast cancer classification from histopathological images with inception recurrent residual convolutional neural network. Journal of digital imaging. 2019 Aug;32(4):605-17.

[54] Sahiner B, Pezeshk A, Hadjiiski LM, Wang X, Drukker K, Cha KH, Summers RM, Giger ML. Deep learning in medical imaging and radiation therapy. Medical physics. 2019 Jan;46(1):e1-36.

[55] Lundervold AS, Lundervold A. An overview of deep learning in medical imaging focusing on MRI. Zeitschrift für Medizinische Physik. 2019 May 1;29(2):102-27.

[56] Huang W, Luo M, Liu X, Zhang P, Ding H, Xue W, Ni D. Arterial spin labeling images synthesis from sMRI using unbalanced deep discriminant learning. IEEE transactions on medical imaging. 2019 Mar 21;38(10):2338-51.

[57] Ghesu FC, Krubasik E, Georgescu B, Singh V, Zheng Y, Hornegger J, Comaniciu D. Marginal space deep learning: efficient architecture for volumetric image parsing. IEEE transactions on medical imaging. 2016 Mar 7;35(5):1217-28.

[58] Krawczyk B, Minku LL, Gama J, Stefanowski J, Woźniak M. Ensemble learning for data stream analysis: A survey. Information Fusion. 2017 Sep 1;37:132-56.

[59] Liu Y, Liu S, Wang Z. Multi-focus image fusion with dense SIFT. Information Fusion. 2015 May 1;23:139-55.

[60] Rahman MA, Zaman N, Asyhari AT, Al-Turjman F, Bhuiyan MZ, Zolkipli MF. Data-driven dynamic clustering framework for mitigating the adverse economic impact of Covid-19 lockdown practices. Sustainable Cities and Society. 2020 Nov 1;62:102372.

[61] Shen D, Wu G, Suk HI. Deep learning in medical image analysis. Annual review of biomedical engineering. 2017 Jun 21;19:221-48.

[62] Litjens G, Kooi T, Bejnordi BE, Setio AA, Ciompi F, Ghafoorian M, Van Der Laak JA, Van Ginneken B, Sánchez CI. A survey on deep learning in medical image analysis. Medical image analysis. 2017 Dec 1;42:60-88.

[63] Ker J, Wang L, Rao J, Lim T. Deep learning applications in medical image analysis. Ieee Access. 2017 Dec 29;6:9375-89.

[64] Gao X, Li W, Loomes M, Wang L. A fused deep learning architecture for viewpoint classification of echocardiography. Information Fusion. 2017 Jul 1;36:103-13.

[65] Lo SC, Lou SL, Lin JS, Freedman MT, Chien MV, Mun SK. Artificial convolution neural network techniques and applications for lung nodule detection. IEEE transactions on medical imaging. 1995 Dec;14(4):711-8.

[66] Shen W, Zhou M, Yang F, Yang C, Tian J. Multi-scale convolutional neural networks for lung nodule classification. InInternational conference on information processing in medical imaging 2015 Jun 28 (pp. 588-599). Springer, Cham.

[67] Rajpurkar P, Irvin J, Zhu K, Yang B, Mehta H, Duan T, Ding D, Bagul A, Langlotz C, Shpanskaya K, Lungren MP. Chexnet: Radiologist-level pneumonia detection on chest x-rays with deep learning. arXiv preprint arXiv:1711.05225. 2017 Nov 14.

[68] Pratt H, Coenen F, Broadbent DM, Harding SP, Zheng Y. Convolutional neural networks for diabetic retinopathy. Procedia computer science. 2016 Jan 1;90:200-5.

[69] Abràmoff MD, Lou Y, Erginay A, Clarida W, Amelon R, Folk JC, Niemeijer M. Improved automated detection of diabetic retinopathy on a publicly available dataset through integration of deep learning. Investigative ophthalmology & visual science. 2016 Oct 1;57(13):5200-6.

[70] Kawahara J, Hamarneh G. Multi-resolution-tract CNN with hybrid pretrained and skin-lesion trained layers. InInternational workshop on machine learning in medical imaging 2016 Oct 17 (pp. 164-171). Springer, Cham.

[71] Roth HR, Lee CT, Shin HC, Seff A, Kim L, Yao J, Lu L, Summers RM. Anatomy-specific classification of medical images using deep convolutional nets. In2015 IEEE 12th international symposium on biomedical imaging (ISBI) 2015 Apr 16 (pp. 101-104). IEEE.

[72] Guo Y, Gao Y, Shen D. Deformable MR prostate segmentation via deep feature learning and sparse patch matching. IEEE transactions on medical imaging. 2015 Dec 11;35(4):1077-89.

[73] Shin HC, Orton MR, Collins DJ, Doran SJ, Leach MO. Stacked autoencoders for unsupervised feature learning and multiple organ detection in a pilot study using 4D patient data. IEEE transactions on pattern analysis and machine intelligence. 2012 Dec 31;35(8):1930-43.

[74] Payer C, Štern D, Bischof H, Urschler M. Regressing heatmaps for multiple landmark localization using CNNs. InInternational Conference on Medical Image Computing and Computer-Assisted Intervention 2016 Oct 17 (pp. 230-238). Springer, Cham.

[75] Baumgartner CF, Kamnitsas K, Matthew J, Smith S, Kainz B, Rueckert D. Real-time standard scan plane detection and localisation in fetal ultrasound using fully convolutional neural networks. InInternational conference on medical image computing and computer-assisted intervention 2016 Oct 17 (pp. 203-211). Springer, Cham.

[76] Duan J, Bello G, Schlemper J, Bai W, Dawes TJ, Biffi C, de Marvao A, Doumoud G, O'Regan DP, Rueckert D. Automatic 3D bi-ventricular segmentation of cardiac images by a shape-refined multi-task deep learning approach. IEEE Transactions on Medical Imaging. 2019 Jan 23;38(9):2151-64.

[77] Albarqouni, S., Baur, C., Achilles, F., Belagiannis, V., Demirci, S. and Navab, N., 2016, "Aggnet: deep learning from crowds for mitosis detection in breast cancer histology images," IEEE Transactions on Medical Imaging, 35(5), pp.1313-1321.

[78] Ghesu FC, Krubasik E, Georgescu B, Singh V, Zheng Y, Hornegger J, Comaniciu D. Marginal space deep learning: efficient architecture for volumetric image parsing. IEEE transactions on medical imaging. 2016 Mar 7;35(5):1217-28.

[79] Shin HC, Roth HR, Gao M, Lu L, Xu Z, Nogues I, Yao J, Mollura D, Summers RM. Deep convolutional neural networks for computer-aided detection: CNN architectures, dataset characteristics and transfer learning. IEEE Transactions on Medical Imaging. 2016 Feb 11;35(5):1285-98.

[80] Peng X, Schmid C. Multi-region two-stream R-CNN for action detection. InEuropean conference on computer vision 2016 Oct 8 (pp. 744-759). Springer, Cham.

[81] Liao F, Liang M, Li Z, Hu X, Song S. Evaluate the malignancy of pulmonary nodules using the 3-d deep leaky noisy-or network. IEEE transactions on neural networks and learning systems. 2019 Feb 14;30(11):3484-95.

[82] Xu J, Xiang L, Liu Q, Gilmore H, Wu J, Tang J, Madabhushi A. Stacked sparse autoencoder (SSAE) for nuclei detection on breast cancer histopathology images. IEEE transactions on medical imaging. 2015 Jul 20;35(1):119-30.

[83]  Cruz-Roa AA, Ovalle JE, Madabhushi A, Osorio FA. A deep learning architecture for image representation, visual interpretability and automated basal-cell carcinoma cancer detection. InInternational Conference on Medical Image Computing and Computer-Assisted Intervention 2013 Sep 22 (pp. 403-410). Springer, Berlin, Heidelberg.

[84]  Khamparia, A., Bharati, S., Podder, P., Gupta, D., Khanna, A., Phung, T.K. and Thanh, D.N., 2021, "Diagnosis of breast cancer based on modern mammography using hybrid transfer learning," Multidimensional Systems and Signal Processing, 32(2), pp.747-765.

[85]  Gu, Z., Cheng, J., Fu, H., Zhou, K., Hao, H., Zhao, Y., Zhang, T., Gao, S. and Liu, J., 2019, "Ce-net: Context encoder network for 2d medical image segmentation," IEEE transactions on medical imaging, 38(10), pp.2281-2292.

[86]  Guo Z, Li X, Huang H, Guo N, Li Q. Deep learning-based image segmentation on multimodal medical imaging. IEEE Transactions on Radiation and Plasma Medical Sciences. 2019 Jan 1;3(2):162-9.

[87]  Oktay O, Ferrante E, Kamnitsas K, Heinrich M, Bai W, Caballero J, Cook SA, De Marvao A, Dawes T, O'Regan DP, Kainz B. Anatomically constrained neural networks (ACNNs): application to cardiac image enhancement and segmentation. IEEE transactions on medical imaging. 2017 Sep 26;37(2):384-95.

[88]  Seebock P, Orlando JI, Schlegl T, Waldstein SM, Bogunović H, Klimscha S, Langs G, Schmidt-Erfurth U. Exploiting epistemic uncertainty of anatomy segmentation for anomaly detection in retinal OCT. IEEE transactions on medical imaging. 2019 May 31;39(1):87-98.

[89]  Zhu, Q., Du, B. and Yan, P., 2019, "Boundary-weighted domain adaptive neural network for prostate MR image segmentation," IEEE transactions on medical imaging, 39(3), pp.753-763.

[90]  Haskins G, Kruger U, Yan P. Deep learning in medical image registration: a survey. Machine Vision and Applications. 2020 Feb;31(1):1-8.

[91]  Elmahdy MS, Jagt T, Zinkstok RT, Qiao Y, Shahzad R, Sokooti H, Yousefi S, Incrocci L, Marijnen CA, Hoogeman M, Staring M. Robust contour propagation using deep learning and image registration for online adaptive proton therapy of prostate cancer. Medical physics. 2019 Aug;46(8):3329-43.

[92]  Bharati, S., Podder, P., Mondal, R., Mahmood, A. and Raihan-Al-Masud, M., 2018, December, "Comparative performance analysis of different classification algorithm for the purpose of prediction of lung cancer," In International Conference on Intelligent Systems Design and Applications, Springer, Cham, pp. 447-457.

[93] Bharati, S., Podder, P. and Paul, P.K., 2019, "Lung cancer recognition and prediction according to random forest ensemble and RUSBoost algorithm using LIDC data," International Journal of Hybrid Intelligent Systems, 15(2), pp.91-100.

[94] Bharati S, Rahman MA, Podder P. Breast cancer prediction applying different classification algorithm with comparative analysis using WEKA. In2018 4th International Conference on Electrical Engineering and Information & Communication Technology (iCEEiCT) 2018 Sep 13 (pp. 581-584). IEEE.

[95] Podder, P., Bharati, S. and Mondal, M.R.H., 2021, "10 Automated gastric cancer detection and classification using machine learning," In Artificial Intelligence for Data-Driven Medical Diagnosis, De Gruyter, pp. 207-224.

[96]  Bai W, Shi W, O'regan DP, Tong T, Wang H, Jamil-Copley S, Peters NS, Rueckert D. A probabilistic patch-based label fusion model for multi-atlas segmentation with registration refinement: application to cardiac MR images. IEEE transactions on medical imaging. 2013 Apr 5;32(7):1302-15.

[97]  Chee E, Wu Z. Airnet: Self-supervised affine registration for 3d medical images using neural networks. arXiv preprint arXiv:1810.02583. 2018 Oct 5.

[98]  Lv J, Yang M, Zhang J, Wang X. Respiratory motion correction for free-breathing 3D abdominal MRI using CNN-based image registration: a feasibility study. The British journal of radiology. 2018 Jan;91(xxxx):20170788.

[99]  Husnayain A, Fuad A, Su EC. Applications of Google Search Trends for risk communication in infectious disease management: A case study of the COVID-19 outbreak in Taiwan. International Journal of Infectious Diseases. 2020 Jun 1;95:221-3.

[100]  Ismael, A. M., & Şengür, A. (2020). Deep Learning Approaches for COVID-19 Detection Based on Chest X-ray Images. Expert Systems with Applications, 114054. doi:10.1016/j.eswa.2020.114054

[101]  Khan, A.I.; Shah, J.L.; Bhat, M.M. CoroNet: A deep neural network for detection and diagnosis of COVID-19 from chest X-ray images. Comput. Methods Programs Biomed. 2020, 196, 105581.

[102]  Brunese, L.; Mercaldo, F.; Reginelli, A.; Santone, A. Explainable deep learning for pulmonary disease and coronavirus COVID-19 detection from X-rays. Comput. Methods Progr. Biomed. 2020, 196, 105608.

[103]  Halgurd S. Maghdid, Aras T. Asaad, Kayhan Zrar Ghafoor, Ali Safaa Sadiq, Seyedali Mirjalili, and Muhammad Khurram Khan "Diagnosing COVID-19 pneumonia from x-ray and CT images using deep learning and transfer learning algorithms", Proc. SPIE 11734, Multimodal Image Exploitation and Learning 2021, 117340E (12 April 2021); https://doi.org/10.1117/12.2588672

[104]  Wang, L.; Lin, Z.Q.; Wong, A. Covid-net: A tailored deep convolutional neural network design for detection of covid-19 cases from chest X-ray images. Sci. Rep. 2020, 10, 1–12.

[105]  Sethy, Prabira & Santi, Kumari & Behera, & Kumar, Pradyumna & Biswas, Preesat. (2020). Detection of coronavirus Disease (COVID-19) based on Deep Features and Support Vector Machine. International Journal of Mathematical, Engineering and Management Sciences, Vol. 5, No. 4, 643-651. doi: 10.33889/IJMEMS.2020.5.4.052

[106]  Gupta, A., Gupta, S., & Katarya, R., 2021, "InstaCovNet-19: A deep learning classification model for the detection of COVID-19 patients using Chest X-ray," Applied Soft Computing, 99, 106859.

[107]  Dhiman, G., Chang, V., Kant Singh, K., & Shankar, A., 2021, "Adopt: automatic deep learning and

optimization-based approach for detection of novel coronavirus covid-19 disease using x-ray images," Journal of biomolecular structure and dynamics, 1-13.

[108] Wang, S., Kang, B., Ma, J. et al., 2021, "A deep learning algorithm using CT images to screen for Corona virus disease (COVID-19)," Eur Radiol, https://doi.org/10.1007/s00330-021-07715-1

[109] Song, Y., Zheng, S., Li, L., Zhang, X., Zhang, X., Huang, Z., Chen, J., Wang, R., Zhao, H., Zha, Y. and Shen, J., 2021. Deep learning enables accurate diagnosis of novel coronavirus (COVID-19) with CT images. IEEE/ACM Transactions on Computational Biology and Bioinformatics.

[110] Shah, V., Keniya, R., Shridharani, A. et al. Diagnosis of COVID-19 using CT scan images and deep learning techniques. Emerg Radiol (2021). https://doi.org/10.1007/s10140-020-01886-y

[111] Tao Zhou, Huiling Lu, Zaoli Yang, Shi Qiu, Bingqiang Huo, Yali Dong,"The ensemble deep learning model for novel COVID-19 on CT images", Applied Soft Computing, Volume 98, 2021, 106885. https://doi.org/10.1016/j.asoc.2020.106885

[112] Mohammad Rahimzadeh, Abolfazl Attar, Seyed Mohammad Sakhaei, "A fully automated deep learning-based network for detecting COVID-19 from a new and large lung CT scan dataset", Biomedical Signal Processing and Control, Volume 68, 2021, 102588. https://doi.org/10.1016/j.bspc.2021.102588.

[113] Shalbaf, A. and Vafaeezadeh, M., 2021, "Automated detection of COVID-19 using ensemble of transfer learning with deep convolutional neural network based on CT scans," International journal of computer assisted radiology and surgery, 16(1), pp.115-123.

[114] Chen, B., et al.: Predicting HLA class II antigen presentation through integrated deep learning. Nat. Biotechnol. 37(11), 1332–1343 (2019)

[115] Kim, Y., et al.: A computational framework for deep learning-based epitope prediction by using structure and sequence information. In: 2019 IEEE International Conference on Bioinformatics and Biomedicine (BIBM), pp. 1208–1210. IEEE (2019)

[116] Zhu, R.-F., Gao, R.-L., Robert, S.-H., Gao, J.-P., Yang, S.-G., Zhu, C.: Systematic review of the registered clinical trials of coronavirus diseases 2019 (COVID-19). medRxiv (2020)

[117] Díaz, Ó., Dalton, J.A., Giraldo, J.: Artificial intelligence: a novel approach for drug discovery. Trends Pharmacol. Sci. 40(8), 550–551 (2019)

[118] Jamshidi, M.B., Jamshidi, M., Rostami, S.: An intelligent approach for nonlinear system identification of a li-ion battery. In: 2017 IEEE 2nd International Conference on Automatic Control and Intelligent Systems (I2CACIS), pp. 98–103. IEEE (2017)

[119] Chen, H., Engkvist, O., Wang, Y., Olivecrona, M., Blaschke, T.: The rise of deep learning in drug discovery. Drug Discov. Today 23(6), 1241–1250 (2018)

[120] Ghasemi, F., Mehridehnavi, A., Perez-Garrido, A., Perez-Sanchez, H.: Neural network and deep-learning algorithms used in QSAR studies: merits and drawbacks. Drug Discov. Today (10), 1784–1790 (2018)

[121] Sunseri, J., King, J.E., Francoeur, P.G., Koes, D.R.: Convolutional neural network scoring and minimization in the D3R 2017 community challenge. J. Comput.-Aided Mol. Des. 33(1), 19–34 (2019)

[122] Gawehn, E., Hiss, J.A., Schneider, G.: Deep learning in drug discovery. Mol. Inform. 35(1), 3–14 (2016)

[123] Hinton, G., et al.: Deep neural networks for acoustic modeling in speech recognition: the shared views of four research groups. IEEE Sig. Process. Mag. 29(6), 82–97 (2012)

[124] Jing, Y., Bian, Y., Hu, Z., Wang, L., Xie, X.-Q.S.: Deep learning for drug design: an artificial intelligence paradigm for drug discovery in the big data era. AAPS J 20(3), 58 (2018)

[125] Sellwood, M.A., Ahmed, M., Segler, M.H., Brown, N.: Artificial intelligence in drug discovery. Future Science (2018)

[126] Fleming, N.: How artificial intelligence is changing drug discovery. Nature 557(7706), S55–S55 (2018)

[127] Larranaga, P., et al.: Machine learning in bioinformatics. Brief. Bioinform. 7(1), 86–112 (2006)

[128] Cherkasov, A., et al.: QSAR modeling: where have you been? where are you going to? J. Med. Chem. 57(12), 4977–5010 (2014)

[129] Smalley, E.: AI-powered drug discovery captures pharma interest. Nat. Biotechnol. 35, 604–605 (2017)

[130] Ye, D.: Artificial intelligence and deep learning application in evaluating the descendants of Tubo Mgar Stong Btsan and social development. In: Data Processing Techniques and Applications for Cyber-Physical Systems (DPTA 2019), pp. 1869–1876. Springer (2020)

[131] Xu, B., et al.: Epidemiological data from the COVID-19 outbreak, real-time case information. Sci. Data 7(1), 1–6 (2020)

[132] Dong, E., Du, H., Gardner, L.: An interactive web-based dashboard to track COVID-19 in real time. Lancet. Infect. Dis. 20(5), 533–534 (2020)

[133] Wang, S., He, M., Gao, Z., He, S., Ji, Q.: Emotion recognition from thermal infrared images using deep Boltzmann machine. Front. Comput. Sci. 8(4), 609–618 (2014)

[134] Salakhutdinov, R., Hinton, G.: Deep Boltzmann machines. In: Artificial Intelligence and Statistics, pp. 448–455 (2009)

[135] Taherkhani, A., Cosma, G., McGinnity, T.M.: Deep-FS: a feature selection algorithm for deep Boltzmann machines. Neurocomputing 322, 22–37 (2018).

[136] Alberici, D., Barra, A., Contucci, P., Mingione, E.: Annealing and replica-symmetry in deep Boltzmann machines. J. Stat. Phys., 1–13 (2020)

[137] Hess, M., Lenz, S., Blätte, T.J., Bullinger, L., Binder, H.: Partitioned learning of deep Boltzmann machines for SNP data. Bioinformatics 33(20), 3173–3180 (2017)

[138] Zhang, X.-L., Wu, J.: Deep belief networks based voice activity detection. IEEE Trans. Audio Speech Lang. Process. 21(4), 697–710 (2012).

[139] Hinton, G.E., Osindero, S., Teh, Y.-W.: A fast learning algorithm for deep belief nets. Neural Comput. 18(7), 1527–1554 (2006).

[140] Hinton, G.E., Salakhutdinov, R.R.: Reducing the dimensionality of data with neural networks. Science 313(5786), 504–507 (2006).

[141] Maetschke, S.R., Ragan, M.A.J.B.: Characterizing cancer subtypes as attractors of Hopfield networks 30(9), 1273–1279 (2014)

[142]  Conforte, A.J., Alves, L., Coelho, F.C., Carels, N., Silva, F.A.B.D.: Modeling basins of attraction for breast cancer using Hopfield networks. Front. Genet. 11, 314 (2020)

[143]  Al-Maitah, M.: Analyzing genetic diseases using multimedia processing techniques associative decision tree-based learning and Hopfield dynamic neural networks from medical images. Neural Comput. Appl. 32(3), 791–803 (2020)

[144]  Stephenson, N., et al.: Survey of machine learning techniques in drug discovery. Curr. Drug Metab. 20(3), 185–193 (2019)

[145]  Rifaioglu, A.S., Atas, H., Martin, M.J., Cetin-Atalay, R., Atalay, V., Doğan, T.: Recent applications of deep learning and machine intelligence in silico drug discovery: methods, tools and databases. Brief. Bioinform. 20(5), 1878–1912 (2019)

[146]  Sangari, A., Sethares, W.: Paper texture classification via multi-scale restricted Boltzman machines. In: 2014 48th Asilomar Conference on Signals, Systems and Computers, pp. 482–486. IEEE (2014)

[147]  Kim, H.-C., Jang, H., Lee, J.-H.: Test–retest reliability of spatial patterns from resting-state functional MRI using the restricted Boltzmann machine and hierarchically organized spatial patterns from the deep belief network. J. Neurosci. Meth. 330, 108451 (2020).

[148]  Smolensky, P.: Information processing in dynamical systems: foundations of harmony theory. Department of Computer Science, Colorado University at Boulder (1986).

[149]  Cai, C., et al.: Transfer learning for drug discovery. J. Med. Chem. 63(16), 8683–8694 (2020)

[150]  Rao, P.M.M., Singh, S.K., Khamparia, A., Bhushan, B. and Podder, P., 2021, "Multi-class Breast Cancer Classification using Ensemble of Pretrained models and Transfer Learning", Current Medical Imaging.

[151]  Karim, M.R., Cochez, M., Jares, J.B., Uddin, M., Beyan, O., Decker, S.: Drug-drug interaction prediction based on knowledge graph embeddings and convolutional-LSTM network. In: Proceedings of the 10th ACM International Conference on Bioinformatics, Computational Biology and Health Informatics, pp. 113–123 (2019)

[152]  Conover, M., Staples, M., Si, D., Sun, M., Cao, R.: AngularQA: protein model quality assessment with LSTM networks. Comput. Math. Biophys. 7(1), 1–9 (2019).

[153]  Narin A, Kaya C, Pamuk Z. Automatic detection of coronavirus disease (covid-19) using x-ray images and deep convolutional neural networks. Pattern Analysis and Applications. 2021 May 9:1-4.

[154]  Minaee S, Kafieh R, Sonka M, Yazdani S, Soufi GJ. Deep-covid: Predicting covid-19 from chest x-ray images using deep transfer learning. Medical image analysis. 2020 Oct 1;65:101794.

[155]  Abbas A, Abdelsamea MM, Gaber MM. Classification of COVID-19 in chest X-ray images using DeTraC deep convolutional neural network. Applied Intelligence. 2021 Feb;51(2):854-64.

[156]  Harmon SA, Sanford TH, Xu S, Turkbey EB, Roth H, Xu Z, Yang D, Myronenko A, Anderson V, Amalou A, Blain M. Artificial intelligence for the detection of COVID-19 pneumonia on chest CT using multinational

datasets. Nature communications. 2020 Aug 14;11(1):1-7.

[157]  Mukherjee H, Ghosh S, Dhar A, Obaidullah SM, Santosh KC, Roy K. Deep neural network to detect COVID-19: one architecture for both CT Scans and Chest X-rays. Applied Intelligence. 2021 May;51(5):2777-89.

[158]  Podder, P., Bharati, S., Mondal, M.R.H. and Kose, U., 2021, "Application of Machine Learning for the Diagnosis of COVID-19. In Data Science for COVID-19," Academic Press, pp. 175-194.

[159]  Bharati S, Podder P, Mondal MR. Diagnosis of Polycystic Ovary Syndrome Using Machine Learning Algorithms", In 2020 IEEE Region 10 Symposium (TENSYMP) 2020 Jun 5 (pp. 1486-1489). IEEE.

[160]  Roberts, M., Driggs, D., Thorpe, M., Gilbey, J., Yeung, M., Ursprung, S., Aviles-Rivero, A.I., Etmann, C., McCague, C., Beer, L. and Weir-McCall, J.R., 2021. "Common pitfalls and recommendations for using machine learning to detect and prognosticate for COVID-19 using chest radiographs and CT scans", Nature Machine Intelligence, 3(3), pp.199-217.

## Author Biographies

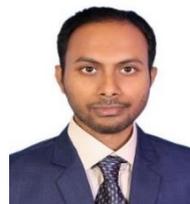**Subrato Bharati** received his B.S. degree in Electrical and Electronic Engineering from Ranada Prasad Shaha University, Narayanganj-1400, Bangladesh. He is currently working as a research assistant in the Institute of Information and Communication Technology(IICT), Bangladesh University of Engineering and Technology, Dhaka, Bangladesh. He is a regular reviewer of some Elsevier and Springer International Journals. He is an associate editor of Journal of the International Academy for Case Studies. He is a member of scientific and technical program committee in some conferences including CECNet 2021, ICONCS, ICCRDA-2020, ICICCR 2021, etc. His research interest includes bioinformatics, medical image processing, pattern recognition, deep learning, wireless communications, data analytics, machine learning, neural networks, and feature selection. He published several journals paper, and also published several IEEE, Springer reputed conference papers. He published Springer and Elsevier, De Gruyter, CRC Press and Wiley Book chapters as well.

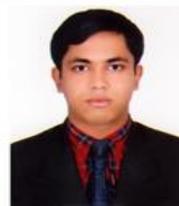**Prajoy Podder** worked as a Lecturer in the Department of Electrical and Electronic Engineering in Ranada Prasad Shaha University, Narayanganj-1400, Bangladesh. He completed B.Sc. (Engg.) degree in Electronics and Communication Engineering from Khulna University of Engineering & Technology, Khulna-9203, Bangladesh. He is currently pursuing M.Sc. (Engg.) degree in Institute of Information and Communication Technology from Bangladesh University of Engineering and Technology, Dhaka-1000, Bangladesh. He is a researcher in the Institute of Information and Communication Technology, Bangladesh University of Engineering & Technology, Dhaka-1000, Bangladesh. He is regular reviewer of Data in Brief, Elsevier and Frontiers of Information Technology and Electronic Engineering, Springer, ARRAY, Elsevier. He is the lead guest editor of Special Issue on Development of Advanced Wireless Communications, Networks and Sensors in American Journal of Networks and Communications. His research interest includes machine learning, pattern recognition, neural networks, computer networking, distributed sensor networks, parallel and distributed computing, VLSI system design, image processing, embedded system design, data analytics. He published several IEEE conference papers, journals and Springer Book Chapters.

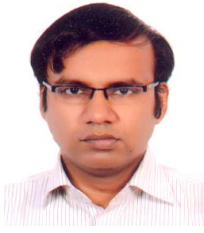

**M. Rubaiyat Hossain Mondal**, PhD is currently working as a faculty member in the Institute of Information and Communication Technology (IICT) at Bangladesh University of Engineering and Technology (BUET), Bangladesh. He received his Bachelor's degree and Master's degree in Electrical and Electronic Engineering from BUET. He joined IICT, BUET as a faculty member in 2005. From 2010 to 2014 he was with the Department of Electrical and Computer Systems Engineering (ECSE) of Monash University, Australia from where he obtained his PhD in 2014. He has authored a number of articles in reputed journals, conferences and book chapters. He is an active reviewer of several journals published by IEEE, Elsevier and Springer. He was a member of the Technical Committee of different IEEE International conferences. His research interest includes artificial intelligence, bioinformatics, image processing, wireless communication and optical wireless communication.

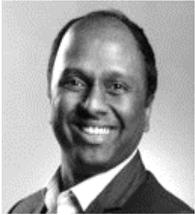

**V. B. Surya Prasath** graduated from the Indian Institute of Technology Madras, India in 2009 with a PhD in Mathematics. He is currently an assistant professor in the Division of Biomedical Informatics at the Cincinnati Children's Hospital Medical Center, and at the Departments of Biomedical Informatics, Electrical Engineering and Computer Science, University of Cincinnati from 2018. He has been a postdoctoral fellow at the Department of Mathematics, University of Coimbra, Portugal, for two years from 2010 to 2011. From 2012 to 2015 he was with the Computational Imaging and VisAnalysis (CIVA) Lab at the University of Missouri, USA as a postdoctoral fellow, and from 2016 to 2017 as an assistant research professor. He had summer fellowships/visits at Kitware Inc. NY, USA, The Fields Institute, Canada, and IPAM, University of California Los Angeles (UCLA), USA. His main research interests include nonlinear PDEs, regularization methods, inverse and ill-posed problems, variational, PDE based image processing, and computer vision with applications in remote sensing, biomedical imaging domains. His current research focuses are in data science, and bioimage informatics with machine learning techniques.